\documentclass[aps,showpacs,nofootinbib,superscriptaddress]{revtex4}
\usepackage{graphicx}
\usepackage{dcolumn}

\begin{document}

\title{ Study of the leptonic  decays of pseudoscalar $B, D$ and vector
$B^*, D^*$ 
mesons and of the semileptonic $B\to D$ and $B\to D^*$ decays.}
\author{C. Albertus} \affiliation{Departamento de
F\'{\i}sica Moderna, Universidad de Granada, E-18071 Granada, Spain.}
\author{ E. Hern\'andez} \affiliation{Grupo de F\'\i sica Nuclear,
Facultad de Ciencias, E-37008 Salamanca, Spain.}  
\author {J. Nieves}
\affiliation{Departamento de F\'{\i}sica Moderna, Universidad de
Granada, E-18071 Granada, Spain.}
\author{ J. M. Verde-Velasco} \affiliation{Grupo de F\'\i sica Nuclear,
Facultad de Ciencias, E-37008 Salamanca, Spain.} 
\begin{abstract} 
\rule{0ex}{3ex} 

\end{abstract}


\begin{abstract} 
 We present  results for different observables in weak decays of
pseudoscalar and vector mesons with a heavy $c$ or $b$ quark. The calculations
are done in a nonrelativistic constituent quark model improved at some instances
by heavy quark effective theory constraints. We determine  pseudoscalar and vector
meson decay constants that within a few per cent satisfy  $f_V M_V/f_P M_P=1$, a result expected in heavy quark symmetry
when the heavy quark masses tend to infinity. We  also analyze the semileptonic  $B\to D$ and $B\to D^*$ decays for
which we evaluate the different form factors. Here we impose  heavy quark
effective theory constraints among form factors that are not  satisfied by a
direct quark model calculation. The value of the form factors at zero recoil
allows us to determine, by comparison with experimental data, the value of the
$|V_{cb}|$ Cabbibo-Kobayashi-Maskawa matrix element. From the $B\to D$
semileptonic decay we get $|V_{cb}|=0.040\pm0.006$ in perfect agreement with our
previous determination based on  the study of the semileptonic 
$\Lambda_b\to \Lambda_c$ decay  and also in excellent agreement
with a recent experimental determination by the DELPHI Collaboration.  
We further make use of the partial conservation of axial current hypothesis to
determine the  strong coupling constants $g_{B^*B\pi}(0)=60.5\pm 1.1$ and
$g_{D^*D\pi}(0)=22.1\pm0.4$. The ratio $R=(\ g_{B^*B\pi}(0)\,f_{B^*}\sqrt{M_D}\
)/
(\ g_{D^*D\pi}(0)\,f_{D^*}\sqrt{M_B}\ )=1.105\pm0.005$ agrees with the heavy quark
symmetry prediction of 1.
\end{abstract}
\maketitle

\section{Introduction}

In systems with a heavy quark with   mass much larger than the QCD scale
($\Lambda_{QCD}$) a new symmetry, known as heavy quark symmetry
 (HQS)~\cite{nussinov,voloshin,politzer,iw}, arises.
In that limit the dynamics of the light quark degrees of freedom becomes
independent of the heavy quark flavor and spin. This is similar to what happens
in atomic physics where the electron properties are approximately independent of
the  spin and mass of the nucleus (for a fixed nuclear charge).
HQS can be cast into the language of an effective theory (HQET)\cite{georgi} 
that allows a 
systematic, order by order, evaluation of  corrections to the infinity mass limit in 
inverse powers of the heavy quark masses. HQS and  HQET 
 have proved  very useful tools to understand bottom
and charm physics and they have been extensively used to
describe the dynamics of systems containing a heavy $c$ or $b$ 
quark~\cite{neubertreport,korner}. For instance, all lattice QCD simulations
rely on HQS to describe bottom systems~\cite{ukqcdb}.

In a recent publication~\cite{albertus05} we have studied the 
$\Lambda_b^0\to \Lambda^+_c\,l^-\bar{\nu}_l$ and 
$\Xi_b^0\to \Xi^+_c\,l^-\bar{\nu}_l$ reactions in a nonrelativistic quark
 model. The detailed analysis of the different form factors showed how
a direct nonrelativistic calculation does not meet HQET constraints and we had 
to improve our model imposing  HQET relations among  form
factors. Our calculation allowed for a
determination of the $|V_{cb}|$ Cabibbo-Kobayashi-Maskawa (CKM) matrix element
 given by $|V_{cb}|=0.040\pm0.005^{+0.001}_{-0.002}$ in good 
 agreement with a recent determination by the DELPHI Collaboration
 $|V_{cb}|=0.0414\pm0.0012\pm0.0021\pm0.0018$~\cite{delphi04}. What we
 intend to do here is a  study of different weak observables of pseudoscalar and
 vector mesons with a heavy $c$ or
 $b$ quark in a  non relativistic quark model improved at some points
 with HQET constraints.  Weak observables are of great interest as they help
 to probe the quark structure of hadrons and provide information to measure the
 CKM matrix elements.
 
In the case of mesons with a heavy quark HQS leads to many model independent
predictions. For instance in the HQS limit the masses of  the lowest lying (s-wave) 
pseudoscalar and vector mesons with a heavy quark  are degenerate. 
Nonrelativistic quark models  satisfy this constraint: the reduced mass of the
system is just the mass of the light quark and the spin-spin 
terms, that distinguishes vector from pseudoscalar, are zero if the mass of
 the heavy quark goes to infinity. HQS also predicts that the
  masses and leptonic decay constants of pseudoscalars and vector mesons
 are related via $f_P\,M_P=f_V\,M_V$, relation that is also satisfied in the quark model in the HQS limit.
 If one looks now at the form factors for the semileptonic $B\to D$ and $B\to
 D^*$  decays HQS predicts  relations among different form factors
 that are also met by the quark model in the HQS limit. The question is
 to what extent the deviations from the HQS limit evaluated in the quark model
 agree with the constraints deduced from HQET. In addition we will make use of
 these HQET constraints to improve the quark model results and thus come up with
 reliable predictions.

The paper is organized as follows: in section~\ref{sect:wf} we introduce
 the meson wave
functions and interquark potentials we shall use in this work. In section~\ref{sect:leptonic} we analyze the leptonic
decays of pseudoscalar and vector  B and D mesons, determining the different 
decay
constants. In section~\ref{sect:semileptonic} we study the form factors for the
semileptonic $B\to D\,l\,\bar{\nu}$ and $B\to D^*\,l\,\bar{\nu}$ decays. In 
section~\ref{sect:gh*hpi} we evaluate the strong coupling constants
$g_{B^*B\pi}$ and $g_{D^*D\pi}$. Finally in section~\ref{sect:conclusions} we
end with the conclusions. The paper also contains three appendices where we
collect the expressions for the matrix elements that are needed for the evaluation
of different observables.

Apart from lattice QCD and QCD sum rules (QCDSR) calculations with which we
shall compare our results and that will be quoted in the following, the different observables analyzed in this work 
have been studied in
the quark model starting with the pioneering work of
 Ref.~\cite{scora95} within a non relativistic version, to continue with
 different versions of the relativistic quark model  applied to the
 determination of decay constants~\cite{capstick90,barik93,hwang96,hwang96b,
 micu97,elhady98,morenas98,melikhov00,wang04,
devito05}, form factors and differential decay widths~\cite{melikhov00,devito05,jaus90,faustov95,barik96,ishida97,
ebert97,ivanov00}, Isgur-Wise functions~\cite{devito05,close94,kiselev95,
leyaouanc96,morenas97,deandrea98,choi99,krutov01} or strong coupling 
constants~\cite{melikhov00,deandrea98,miller88,odonnell94,colangelo94,becirevic99}
\footnote{The list of references is by no means exhaustive.}. 

\section{Wave function and interquark interactions}
\label{sect:wf}
For a meson $M$ we use the following expression for the wave function
\begin{eqnarray}
\label{wf}
&&\hspace{-1cm}\left|{M,\lambda\,\vec{P}}\,\right\rangle_{NR}
=\int d^3p \sum_{\alpha1,\alpha_2}\hat{\phi}^{(M,\lambda)}_{\alpha1,\alpha_2}(\,\vec{p}\,)
\nonumber\\ &&\frac{(-1)^{\frac{1}{2}-s_2}}{(2\pi)^{\frac{3}{2}}
\sqrt{2E_{f_1}(\vec{p}_1)2E_{f_2}(\vec{p}_2)}}\ 
\left|\ q,\ \alpha_1\
\vec{p}_1=\frac{m_{f_1}}{m_{f_1}+m_{f_2}}\vec{P}-\vec{p}\ \right\rangle
\left|\ \bar{q},\ \alpha_2\ \vec{p}_2=\frac{m_{f_2}}{m_{f_1}+m_{f_2}}\vec{P}+\vec{p}\
 \right\rangle
\end{eqnarray}
where $\vec{P}$ stands for the meson three momentum and $\lambda$ represents
the spin projection in the meson center of mass. $\alpha_1$ and $\alpha_2$ represent
the quantum numbers of spin (s), flavor (f) and color (c)
\begin{equation}
\alpha\equiv(s,f,c)
\end{equation}
of the
quark and the antiquark, while $E_{f_1},\,\vec{p}_1$ and $E_{f_2},\,\vec{p}_2$ are their 
respective energies and three-momenta. $m_f$ is the mass of the quark or
antiquark with flavor $f$. The factor $(-1)^{\frac{1}{2}-s_2}$ 
is included in order that the antiquark spin states have the correct relative
phase\footnote{Note that under charge conjugation $({\cal C})$  quark and
antiquark creation operators are related via \hbox{${\cal
C}\,c^{\dagger}_\alpha(\,\vec{p}\,)\,{\cal C}^{\dagger}= (-1)^{\frac{1}{2}-s}\,d^{\dagger}_\alpha(\,\vec{p}\,)
$}. This means that the antiquark states with the correct spin relative phase 
are not $d^{\dagger}_\alpha(\,\vec{p}\,)\,|0\rangle=
|\,\bar{q},\ \alpha\ \vec{p}\ \rangle$ but are instead given by
$(-1)^{\frac{1}{2}-s}\,d^{\dagger}_\alpha(\,\vec{p}\,)\,|0\rangle=
(-1)^{\frac{1}{2}-s}\,|\,\bar{q},\ \alpha\ \vec{p}\ \rangle$.}.
The normalization of the quark and antiquark
states is 
\begin{eqnarray}
\left\langle\ \alpha^{\prime}\ \vec{p}^{\ \prime}\,|\,\alpha\ \vec{p}\,
\right\rangle=\delta_{\alpha^{\prime},\ \alpha}\, (2\pi)^3\, 2E\,\delta(
\vec{p}^{\ \prime}-\vec{p}\,)
\end{eqnarray}
Furthermore, $\hat{\phi}^{\,(M,\lambda)}_{\alpha1,\alpha_2}(\,\vec{p}\,)$ is the
momentum space wave
function  for the relative motion of the quark-antiquark
system. 
Its normalization is given  by
\begin{equation}
\int\, d^3p\ \sum_{\alpha_1\,\alpha_2} \left(
\hat{\phi}^{\,(M,\lambda')}_{\alpha_1,\,\alpha_2}(\,\vec{p}\,)
\right)^* \hat{\phi}^{\,(M,\lambda)}_{\alpha_1,\,\alpha_2}(\,\vec{p}\,)
=\delta_{\lambda',\,\lambda}
\end{equation}
and, thus,  the normalization of our meson states is 
\begin{equation}
{}_{\stackrel{}{NR}}\left\langle\, {M,\lambda'\,\vec{P}^{\,\prime}}\,|\,{M,\lambda
\,\vec{P}}\,\right\rangle_{NR}
=\delta_{\lambda',\,\lambda}\,(2\pi)^3\,\delta(\vec{P}^{\,\prime}-\vec{p}\,)
\end{equation}
For the particular case of ground state pseudoscalar ($P$) and vector ($V$) mesons we can assume the
orbital angular momentum to be zero and then we will have
\begin{eqnarray}
\hat{\phi}^{\,(P)}_{\alpha1,\,\alpha_2}(\,\vec{p}\,)
&=&\frac{1}{\sqrt{3}}\,\delta_{c_1,\,c_2}\
\hat{\phi}^{\,(P)}_{(s_1,\,f_1),\,(s_2,\,f_2)}(\,\vec{p}\,)\nonumber\\
&=&\frac{1}{\sqrt{3}}\,\delta_{c_1,\,c_2}\ (-i)
\ \hat{\phi}^{\,(P)}_{f_1,\,f_2}(\,|\vec{p}\,|)\ 
Y_{00}(\,\widehat{\vec{p}}\ )\ (1/2,1/2,0;s_1,s_2,0)\nonumber\\
\hat{\phi}^{\,(V,\,\lambda)}_{\alpha1,\,\alpha_2}(\,\vec{p}\,)
&=&\frac{1}{\sqrt{3}}\,\delta_{c_1,\,c_2}\
\hat{\phi}^{\,(V,\,\lambda)}_{(s_1,\,f_1),\,(s_2,\,f_2)}(\,\vec{p}\,)\nonumber\\
&=&\frac{1}{\sqrt{3}}\,\delta_{c_1,\,c_2}\ (-1)\ 
\hat{\phi}^{\,(V)}_{f_1,\,f_2}(\,|\vec{p}\,|)\
Y_{00}(\,\widehat{\vec{p}}\ )\
(1/2,1/2,1;s_1,s_2,\lambda)
\end{eqnarray}
where $(j_1,j_2,j_3;m_1,m_2,m_3)$ is a Clebsch-Gordan coefficient, 
$Y_{00}=1/\sqrt{4\pi}$ is the $l=m=0$ spherical harmonic, 
 and 
$\hat{\phi}^{\,(M)}_{f_1,\,f_2}(|\,\vec{p}\,|)$ is the
Fourier transform of the radial coordinate space wave function. 
The phases are
introduced for later convenience. 

To evaluate the coordinate space wave function
we shall use several interquark potentials, one suggested by Bhaduri and
collaborators~\cite{bhaduri81} (BHAD), and four suggested by Silvestre-Brac and Semay
~\cite{silvestre96,sbs93} (AL1,AL2,AP1,AP2). The general structure of those
potentials in the quark-antiquark sector is
\begin{equation}
V_{ij}^{q\bar q}(r) = -\frac{\kappa\left ( 1 - e^{-r/r_c}\right )}{r}
+\lambda r^p - \Lambda + \left\{a_0\frac{\kappa}{m_im_j}
\frac{e^{-r/r_0}}{rr_0^2} + \frac{2\pi}{3m_im_j}\kappa^\prime \left (
1 - e^{-r/r_c}\right ) \frac{e^{-r^2/x_0^2}}{\pi^\frac32 x_0^3} \right
\}\vec{\sigma}_i\vec{\sigma}_j  \label{eq:phe}
\end{equation}
with $\vec{\sigma}$ the spin Pauli matrices, $m_i$ the constituent
quark masses  and
\begin{equation}
x_0(m_i,m_j) = A \left ( \frac{2m_im_j}{m_i+m_j} \right )^{-B} 
\label{eq:phebis}
\end{equation}
The potentials considered differ in the form factors used for the
hyperfine terms, the power of the confining term ($p=1$, as suggested
by lattice QCD calculations~\cite{gm84}, or $p=2/3$ which gives the
correct asymptotic Regge trajectories for mesons~\cite{fabre88}), or
the use of a form factor in the one gluon exchange Coulomb potential.
All free parameters in the potentials have been adjusted to reproduce the light ($\pi$,
$\rho$, $K$, $K^*$, etc.) and heavy-light ($D$, $D^*$, $B$, $B^*$,
etc.) meson spectra. They also  lead to precise predictions for the charmed
and bottom baryon ($\Lambda_{c,b}$, $\Sigma_{c,b}$, $\Sigma^*_{c,b}$,
$\Xi_{c,b}$, $\Xi'_{c,b}$, $\Xi^*_{c,b}$, $\Omega_{c,b}$ and
$\Omega_{c,b}^*$) masses~\cite{silvestre96,albertus04} and for the
semileptonic $\Lambda_b^0 \to \Lambda_c^+ l^- {\bar \nu}_l$ and
$\Xi_b^0 \to \Xi_c^+ l^- {\bar \nu}_l$~\cite{albertus05} decays.

We will use the above mentioned interquark interactions to evaluate
the different observables.  This will provide us with a spread of
results that we will consider, and quote, as a theoretical error to
the averaged value that will quote as our central result.

\section{Leptonic decay of pseudoscalar and vector $B$ and $D$ mesons}
\label{sect:leptonic}
In this section we will consider the purely leptonic decay of pseudo scalars 
 ($B$, $D$) and vector  ($B^*$, $D^*$) mesons.  The
 charged weak current operator for a specific pair of quark flavors
$f_1$ and $f_2$ reads
\begin{equation}
J^{f_1\,f_2}_\mu(0)=J^{f_1\,f_2}_{V\,\mu}(0)-J^{f_1\,f_2}_{A\,\mu}(0)
=\sum_{(c_1,\,s_1),\,(c_2,\,s_2)}\,\delta_{c_1,\,c_2}\ \overline{\Psi}_{\alpha_1}(0)\gamma_\mu
(1-\gamma_5)\Psi_{\alpha_2}(0)
\end{equation}
with $\Psi_{\alpha_1}$  a quark field of a definite spin, flavor
and color. The hadronic matrix elements involved in the
 processes can be parametrized in
terms of a unique pseudoscalar $f_P$ or vector $f_V$ decay constant as
\begin{eqnarray}
\label{fpv}
&&\left\langle 0 \right|\, J_\mu^{f_1\,f_2}(0)\,\left|\, P, \vec{P}\,\right\rangle
=\left\langle 0 \right|\, -J_{A\,\mu}^{f_1\,f_2}(0)\,\left|\, P, \vec{P}\,
\right\rangle= -iP_\mu\,f_P \nonumber\\
&&\left\langle 0 \right|\, J_\mu^{f_1\,f_2}(0)\,\left|\, V, \lambda\,\vec{P}\,
\right\rangle
=\left\langle 0 \right|\, J_{V\,\mu}^{f_1\,f_2}(0)\,\left|\, V,\,\lambda\, 
\vec{P}\,
\right\rangle= \varepsilon^{(\lambda)}_\mu(\,\vec{P}\,)M_Vf_V
\end{eqnarray}
where the meson states are normalized such that
\begin{equation}
\left\langle\, {M,\lambda'\,\vec{P}^{\,\prime}}\,|\,{M,\lambda
\,\vec{P}}\,\right\rangle
=\delta_{\lambda',\,\lambda}\,(2\pi)^3\,2E\,\delta(\vec{P}^{\,\prime}-\vec{p}\,)
\end{equation}
In the first of Eqs.(\ref{fpv}) $P_\mu$ is the four-momentum of the meson, while in
the second  $M_V$ and $\varepsilon^{(\lambda)}_\mu(\,\vec{P}\,)$ are the 
mass and the polarization vector of the vector meson. In both cases $f_1$ and $f_2$ are the 
flavors of the quark and the antiquark that make
up the meson. 

Concerns about the experimental determination of the pseudoscalars decay
contants have been raised in Ref.~\cite{burdman95}. There the
effect of radiative decays was analyzed concluding that for $B$ mesons the 
decay constant determination could be greatly affected by radiative 
corrections. In the vector sector, and as rightly
 pointed out in Ref.~\cite{lellouch94}, the vector decay constants are not relevant from a phenomenological point of
view since $B^*$ and $D^*$ will decay through the electromagnetic and/or strong
interaction. They are nevertheless interesting as a mean to test HQS relations.

For  mesons at rest we will obtain
\begin{eqnarray}
&&f_P=\frac{-i}{M_P}\,\left\langle 0 \right|\, J_{A\,0}^{f_1\,f_2}(0)
\,\left|\, P, \vec{0}\,\right\rangle\nonumber\\
&&f_V=\frac{-1}{M_V}\,\left\langle 0 \right|\, J_{V\,3}^{f_1\,f_2}(0)
\,\left|\, V,\,0\, 
\vec{0}\,
\right\rangle
\end{eqnarray}
with $M_P$ the mass of the pseudoscalar meson.
 In our model, and due to the different normalization of our meson states, we shall 
 evaluate the decay constants as
\begin{eqnarray}
&&f_P=-i\,\sqrt{\frac{2}{M_P}}\,\left\langle 0 \right|\, J_{A\,0}^{f_1\,f_2}(0)
\,\left|\, P, \vec{0}\,\right\rangle_{NR}\nonumber\\
&&f_V=-\sqrt{\frac{2}{M_V}}\,\left\langle 0 \right|\, J_{V\,3}^{f_1\,f_2}(0)
\,\left|\, V,\,0\, 
\vec{0}\,
\right\rangle_{NR}
\end{eqnarray}
The  corresponding matrix elements are given in appendix~\ref{app:lep}

The results that we obtain for the different decay constants appear
 in Tables~\ref{tab:fp} and ~\ref{tab:fv}. Starting with $f_D$ and $f_{D_s}$ our
 results are larger that the ones obtained in the lattice by the UKQCD
 Collaboration~\cite{ukqcd01} or the ones evaluated
 using QCD spectral sum rules (QSSR)~\cite{narison}. Not only the independent values are larger but
 also the ratio $f_{D_s}/f_D$ is larger in our case. On the other hand our results
 are in better agreement with other  lattice
 determinations~\cite{simone04,wingate04}. They also  compare very well
 with the experimental measurements of $f_D$ and $f_{D_s}$ in
  Refs.~\cite{cleo04,cleo98,aleph02,opal01,beatrice00,bes05} being our $f_{D_s}/f_D$
 ratio  in very good agreement with the value obtained using recent CLEO Collaboration
 data~\cite{cleo04,cleo98}. As for $f_B$ and $f_{B_s}$, we find a very good
 agreement in the case of $f_{B_s}$ between  our results and 
 the ones obtained in the lattice or with the use of QSSR. For $f_B$ our result
 is smaller and then also our ratio
$f_{B_s}/f_B$ is larger. 
\begin{table}[h!!!!]
\begin{center}
\begin{tabular}{c|ccc}
 & $f_D$ [MeV] & $f_{D_s}$  [MeV]  & $f_{D_s}/f_D$ \\
\hline
& & &\\
This work & $243^{+21}_{-17}$  & $341^{+7}_{-5}$  & $1.41^{+0.08}_{-0.09}$ \\ \hline\\
Experimental data & & & \\
& & &\\
CLEO & $202\pm41\pm17$~\protect\cite{cleo04} &
$280 \pm19\pm28\pm34$~\protect\cite{cleo98} & ---\\
ALEPH~\cite{aleph02} & --- & $285\pm19\pm40$ & ---\\
OPAL~\cite{opal01} & --- & $286\pm44\pm41$ &---\\
BEATRICE~\cite{beatrice00} & --- &$323\pm44\pm 12\pm34$ &--- \\
E653~\cite{e653-96} & ---&$194\pm35\pm20\pm14$ &---\\
BES~\cite{bes05} & $371^{+129}_{-119}\pm 25$ & --- & ---\\ \hline\\
Lattice data & & & \\
& & & \\
UKQCD~\protect\cite{ukqcd01} & $206(4)^{+17}_{-10}$ & $229(3)^{+23}_{-12}$ &
$1.11(1)^{+1}_{-1}$\\
Fermilab Lattice~\cite{simone04} (Preliminary) & $225^{+11}_{-13}\pm21$& 
$263^{+5}_{-9}\pm24$&---\\
M. Wingate {\it et al.}~\cite{wingate04} & ---& $290\pm20\pm29\pm29\pm6$   &---\\
\hline\\
QCD Spectral Sum Rules & & &\\
& &   &\\
S. Narison~\protect\cite{narison} &     $203\pm23$ & $235\pm24$&$1.15\pm0.04$\\
\hline
& & &\\
& & &\\
 & $f_B$ [MeV] & $f_{B_s}$  [MeV]  & $f_{B_s}/f_B$ \\
\hline
& & &\\
This work & $155^{+15}_{-12}$  & $239^{+9}_{-7}$ & $1.54^{+0.09}_{-0.08}$ \\ \hline\\
Lattice data &  & &\\
& & & \\
UKQCD~\protect\cite{ukqcd01} & $195(6)^{+24}_{-23}$& $220(6)^{+23}_{-18}$ &
$1.13(1)^{+1}_{-1}$\\
M. Wingate {\it et al.}~\cite{wingate04} & ---& $260\pm7\pm26\pm8\pm5$   &---\\
Lattice world averages & $200\pm30$~\cite{bernard01} & $230\pm30$
~\cite{hashimoto04} &  $1.16\pm0.04$~\cite{bernard01}\\ \hline\\
QCD Spectral Sum Rules & & &\\
& &   &\\
S. Narison~\protect\cite{narison} & $207\pm21$ & $240\pm24$& $1.16\pm0.04$\\
\end{tabular}
\end{center}
\caption{ Pseudoscalar $f_P$ decay constants  for $B$ and $D$ mesons}\vspace{1cm}
\label{tab:fp}
\end{table}
\begin{table}[h!!!]
\begin{center}
\begin{tabular}{c|ccc}
 & $\tilde{f}_{D^*}$  & $\tilde{f}_{D^*_s}$    &
 $\tilde{f}_{D^*}/\tilde{f}_{D^*_s}$ \\
\hline
& & &\\
This work & $9.1^{+0.9}_{-0.9}$  & $6.5^{+0.3}_{-0.4}$ & $1.41^{+0.06}_{-0.05}$ \\
UKQCD~\cite{ukqcd01} & $8.6(3)^{+5}_{-9}$ & $8.3(2)^{+5}_{-5}$&
$1.04(1)^{+2}_{-2}$\\
\hline
& & &\\
 & $\tilde{f}_{B^*}$ [MeV] & $\tilde{f}_{B^*_s}$   [MeV]  & $\tilde{f}_{B^*}
 /\tilde{f}_{B^*_s}$ \\
\hline
& & &\\
This work & $35.6^{+3.4}_{-3.4}$  & $23.0^{+1.0}_{-1.5}$ & $1.55^{+0.07}_{-0.06}$ \\
UKQCD~\cite{ukqcd01} & $28(1)^{+3}_{-4}$ & $25(1)^{+2}_{-3}$&
$1.10(2)^{+2}_{-2}$\\
\end{tabular}
\end{center}
\caption{ $\tilde{f}_V=M_V/f_V$   for $B^*$ and $D^*$ mesons}
\label{tab:fv}
\end{table}
\noindent

For the vector meson decay constants we obtain the values
\begin{eqnarray}
f_{D^*}=223^{+23}_{-19}\, {\mathrm MeV} \hspace{1cm} f_{D_s^*}=326^{+21}_{-17}\, {\mathrm MeV}\nonumber\\
f_{B^*}=151^{+15}_{-13}\, {\mathrm MeV} \hspace{1cm} f_{B_s^*}=236^{+14}_{-11}\, {\mathrm MeV}
\end{eqnarray}
which are very much the same as the values obtained
for the decay constants of their pseudoscalar counterparts. This almost equality
of pseudoscalar and vector decay constants is expected in HQS  in the limit
where the heavy quark masses go to infinity where one would have~\cite{iw}
\begin{eqnarray}
f_V\,M_V=f_P\,M_P \hspace{1cm},\hspace{1cm} M_V=M_P
\end{eqnarray}
Our decay constants  satisfy the above relation within  2\%. On the
other hand 
UKQCD lattice data show deviations as large as 20\% for D mesons~\cite{ukqcd01}.
 
In order to compare  the values of the vector decay constants with lattice data
 from Ref.~\cite{ukqcd01} we give in 
Table~\ref{tab:fv} the quantity 
 $\tilde{f}_V=M_V/f_V$. We find good agreement  for 
$\tilde{f}_{D^*}$ and
$\tilde{f}_{B_s^*}$ but not so much for the other two. Also our ratios
$\tilde{f}_{D^*}/\tilde{f}_{D_s^*}$ and $\tilde{f}_{B^*}/\tilde{f}_{B_s^*}$
are larger than the ones favored by lattice calculations. 

On the other hand the
ratio
\begin{eqnarray}
\frac{f_{B^*}\sqrt{M_B}}{f_{D^*}\sqrt{M_D}}=1.138^{+0.011}_{-0.008}
\end{eqnarray}
is in very good agreement with the expectation in Ref.~\cite{burdman94} where
they would get
$1.05\sim1.20$ for that ratio.

\section{Semileptonic decay of $B$ into $Dl\bar{\nu}$ and $D^*l\bar{\nu}$}
\label{sect:semileptonic}
In this case the  strong matrix elements are parametrized as
\begin{eqnarray}
\frac{\left\langle\, D,\,\vec{P}^\prime\,\left|\, J^{c\,b}_\mu(0)\,
\right| \, B,\,\vec{P}\right\rangle}{\sqrt{M_B
M_D}}&=&\frac{\left\langle\, D,\,\vec{P}^\prime\,\left|\, J^{c\,b}_{V\,\mu}(0)\,
\right| \, B,\,\vec{P}\right\rangle}{\sqrt{M_B
M_D}}\nonumber\\ &=&
\left(v+v^\prime\right)_\mu\,h_+(w)+\left(v-v^\prime\right)_\mu\,h_-(w)
\end{eqnarray}
\begin{eqnarray}
\label{bd*}
\frac{\left\langle\, D^*,\,\lambda\,\vec{P}^\prime\,\left|\, J^{c\,b}_\mu(0)\,
\right| \, B,\,\vec{P}\right\rangle}{\sqrt{M_B
M_{D^*}}}&=&\varepsilon_{\mu\nu\alpha\beta}\,\left(\varepsilon^{(\lambda)\,
\nu}(\,\vec{P}^\prime\,)\right)^*\,v^\alpha\,v^{\prime\,
\beta}\,h_V(w)\nonumber\\
&&\hspace{-.3cm}-\, i\, \left(\varepsilon^{(\lambda)}_\mu(\,\vec{P}^\prime\,)\right)^*\,(w+1)
\,h_{A_1}(w)\nonumber\\
&&\hspace{-.3cm}+\, i\, \left(\varepsilon^{(\lambda)\,
}(\,\vec{P}^\prime\,)\right)^*\cdot v\,     \left( 
v_\mu\,h_{A_2}(w)+v^\prime_\mu\,h_{A_3}(w)\right)
\end{eqnarray}
where $v=P/M_B$ and $v^\prime=P^\prime/M_{D,\,D^*}$  are
the four velocities of the initial $B$ and final $D,\,D^*$ mesons, 
$w=v\cdot v^\prime$\ \footnote{$w$ is related to the four momentum transferred
 square
$q^2$ via  $q^2=M_B^2+M_{D,\,D^*}^2-2\ w\, M_B\,M_{D,\,D^*}$} and
$\varepsilon_{\mu\nu\alpha\beta}$ is the fully antisymmetric tensor with 
$\varepsilon_{0123}=+1$.

In the limit of infinite heavy quark masses $m_c,\,m_b\to \infty$ 
HQS reduces the six form factors to  a unique universal function
$\xi(w)$ known as the Isgur-Wise function~\cite{iw}
\begin{eqnarray}
&&h_+(w)=h_{A_1}(w)=h_{A_3}(w)=h_{V}(w)=\xi(w)\\
&&h_{-}(w)=h_{A_2}(w)=0
\end{eqnarray}
Vector current conservation in the equal mass case implies the normalization
\begin{equation}
\xi(1)=1
\end{equation}
Away from the heavy quark limit those relations are modified by QCD corrections
so that one has
\begin{equation}
\label{hiw}
h_j(w)=\left( \alpha_j+\beta_j(w)+\gamma_j(w)+{\cal
O}(1/m_{c,b}^2)\right)\,\xi(w)
\end{equation}
The $\alpha_j$ are constants  fixed by the behavior of the form factor in the
heavy quark limit
\begin{eqnarray}
&&\alpha_+=\alpha_{A_1}=\alpha_{A_3}=\alpha_{V}=1\nonumber\\
&&\alpha_{-}=\alpha_{A_2}=0
\end{eqnarray}
The different $\beta_j$ account for perturbative radiative corrections~\cite{neubert921}
while the $\gamma_j$ are non perturbative in nature and are proportional to the inverse
of the heavy quark masses~\cite{neubert922}. At zero recoil ($w=1$) Luke's
theorem~\cite{luke90} imposes the restriction. 
\begin{equation}
\gamma_+(1)=\gamma_{A_1}(1)=0
\end{equation}
so that power corrections to $h_+(1)$ and $h_{A_1}(1)$ are of order 
${\cal O}(1/m_{c,b}^2)$. In Tables~\ref{tab:beta}
and~\ref{tab:gamma} we collect the values for the different $\beta_j$ and $\gamma_j$
in the full interval of $w$ values allowed in the two decays. These two tables have
been taken from Ref.~\cite{neubert922}.
\begin{table}[t]
\begin{center}
\begin{tabular}{c|r r r r r r}
w & $\beta_+$  & $\beta_{-}$    & $\beta_{V}$ & $\beta_{A_1}$ & $\beta_{A_2}$
& $\beta_{A_3}$\\
\hline
1.0 & 2.6  & $-$5.4 & 11.9 & $-$1.5 & $-$11.0 & 2.2 \\
1.1 & $-$0.3 & $-$5.4 & 8.9 & $-$3.8 & $-$10.3 & $-$0.2 \\
1.2 & $-$3.1 & $-$5.3 & 6.1 & $-$5.9 & $-$9.8 & $-$2.5 \\
1.3 & $-$5.6 & $-$5.3 & 3.5 & $-$7.9 & $-$9.3 & $-$4.6 \\
1.4 & $-$8.0 & $-$5.2 & 1.1 & $-$9.7 & $-$8.8 & $-$6.6 \\
1.5 & $-$10.2 & $-$5.2 & $-$1.1 & $-$11.5 & $-$8.4 & $-$8.5 \\
1.59 & $-$12.1 & $-$5.1 & & & &\\
\end{tabular}
\end{center}
\caption{ QCD corrections $\beta_j(w)$ in \% as evaluated in
Ref.~\cite{neubert922}}
\label{tab:beta}
\end{table}
\begin{table}[h!!]
\begin{center}
\begin{tabular}{c|r r r r r r}
w & $\gamma_+$  & $\gamma_{-}$    & $\gamma_{V}$ & $\gamma_{A_1}$ & $\gamma_{A_2}$
& $\gamma_{A_3}$\\
\hline
1.0 & 0.0 & $-$4.1 & 19.1 & 0.0 & $-$23.1 &$-$4.1 \\
1.1 & 2.7 & $-$4.1 & 20.7 & 2.9 & $-$21.4 & $-$0.7  \\
1.2 & 6.2 & $-$4.1 & 23.1 & 6.5 & $-$19.8 & 3.4 \\
1.3 & 10.5 & $-$4.2 & 26.3 & 10.7 & $-$18.3 & 8.0\\
1.4 & 15.3 &  $-$4.4 & 30.0 & 15.4 & $-$17.0 &13.0 \\
1.5 & 20.6 & $-$4.5 & 34.3 & 20.5 & $-$15.8 & 18.5 \\
1.59 & 25.7 & $-$4.7\\
\end{tabular}
\end{center}
\caption{ Power corrections $\gamma_j(w)$ in \% as evaluated in
Ref.~\cite{neubert922}}
\label{tab:gamma}
\end{table}
\subsection{$B\to D\, l\, \bar{\nu}$ decay}
\label{subsect:btod}
Let us start with the $B\to D\, l\, \bar{\nu}$ case. In the center of mass of the $B$ meson and taking $\vec{P}^{\,\prime}=-\vec{q}=-|\vec{q}\,|\,
\widehat{\vec{k}}$ in the z 
direction  we will have for the form factors $h_+(w)$ and $h_-(w)$\footnote{In this case 
$w$ is related to $|\vec{q}\,|$ via $|\vec{q}\,|=M_D\,\sqrt{w^2-1}$}
\begin{eqnarray}
\label{hpm}
h_+(w)=\frac{1}{\sqrt{2M_B2M_D}}\,\left(
V^0(|\vec{q}\,|)+\frac{V^3(|\vec{q}\,|)}{|\vec{q}\,|}\,\left(E_D(|\vec{q}\,|)-
M_D\right)\right)\nonumber\\
h_-(w)=\frac{1}{\sqrt{2M_B2M_D}}\,\left(
V^0(|\vec{q}\,|)+\frac{V^3(|\vec{q}\,|)}{|\vec{q}\,|}\,\left(
E_D(|\vec{q}\,|)+M_D
\right)\right)
\end{eqnarray}
where $E_D(|\vec{q}\,|)=\sqrt{M_D^2+|\vec{q}\,|^2}$ and $V^\mu(|\vec{q}\,|)$ ($\mu=0,3$)
is given by
\begin{eqnarray}
V^{\mu}(|\vec{q}\,|)=\left\langle\, D,\,-|\vec{q}\,|\,\vec{k}\,\left|\, (J^{c\,b}_V)^\mu(0)\,
\right| \, B,\,\vec{0}\right\rangle
\end{eqnarray}
In our model $V^\mu(|\vec{q}\,|)$ is evaluated as
\begin{eqnarray}
\label{vmu}
V^\mu(|\vec{q}\,|)&=&\sqrt{2M_B2E_D(|\vec{q}\,|)}\ \ 
{}_{\stackrel{}{NR}}\left\langle\, D,\,
-|\vec{q}\,|\,\widehat{\vec{k}}\,\left|\, (J^{c\,b}_V)^\mu(0)\,
\right| \, B,\,\vec{0}\right\rangle_{NR}\nonumber\\
\end{eqnarray}
which expression is given in appendix~\ref{app:v}.

In the case of equal masses $m_b=m_c$ vector current conservation demands that
\begin{eqnarray}
h_+(1)=1\hspace{1cm};\hspace{1cm}
h_-(w)=0
\end{eqnarray}
In this limit we find that $h_+(1)=1$ 
so that our value for $h_+(1)$ complies with vector current conservation. On the
other hand $h_-(w)\ne 0$  
violating vector current conservation.
\begin{figure}[t]
\vspace{1cm}
\centering
\resizebox{7cm}{5cm}{\includegraphics{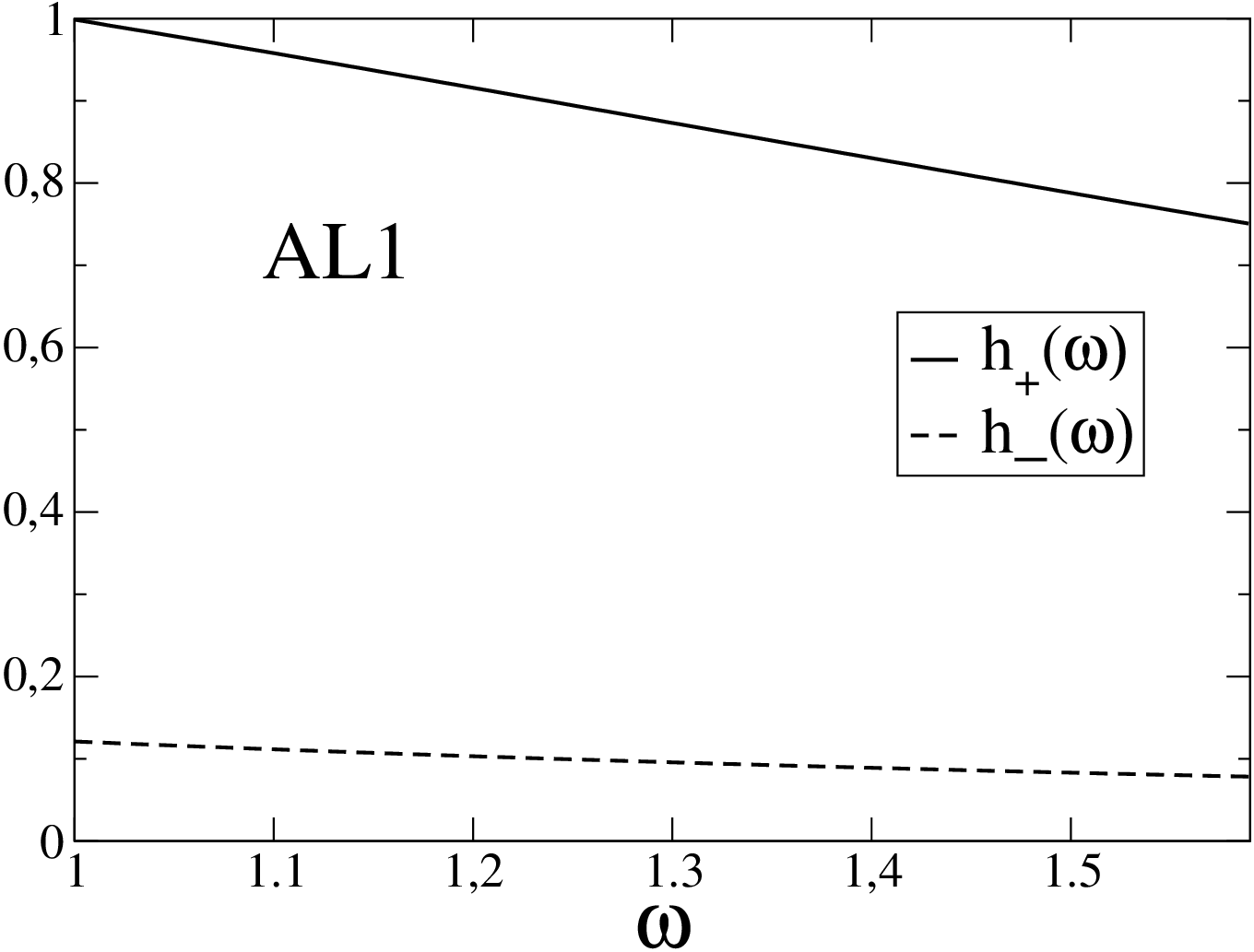}}\hspace{1cm}
\resizebox{7cm}{5cm}{\includegraphics{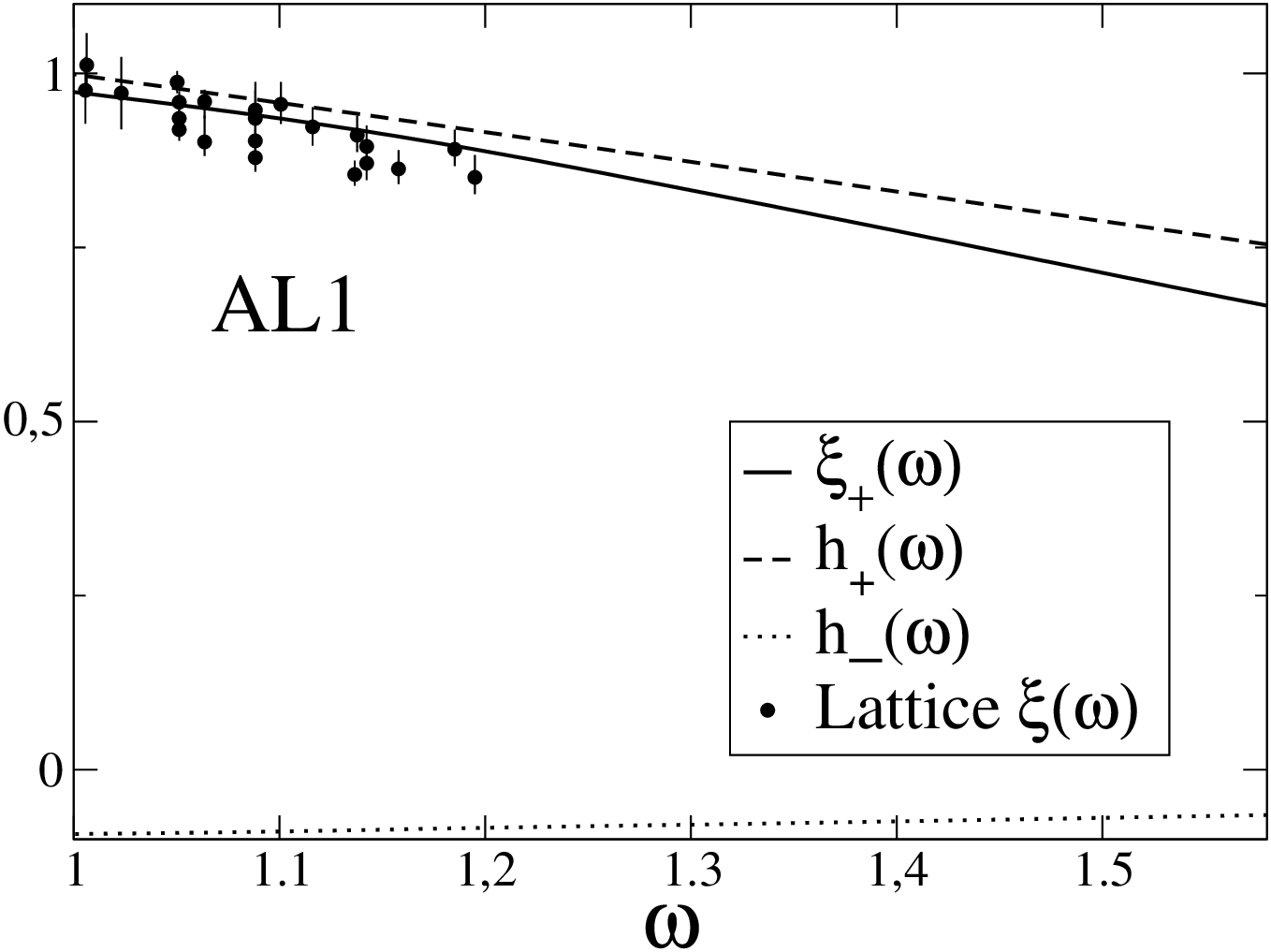}}\vspace{.5cm}\\
\resizebox{7cm}{5cm}{\includegraphics{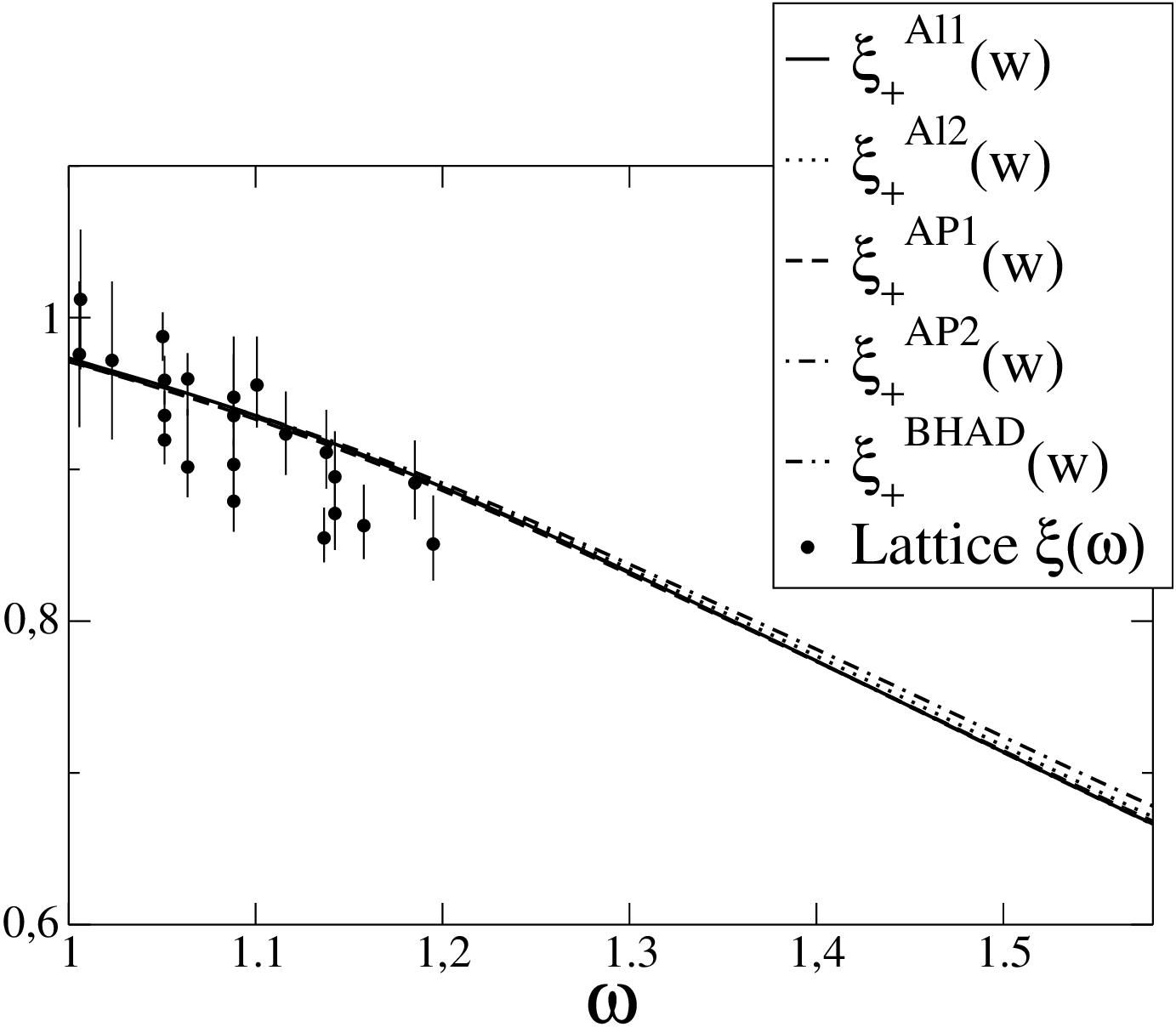}}
\caption{Upper left panel: $h_+(w)$ and $h_-(w)$  for the $B\to D$
transition as obtained from Eqs.~(\ref{hpm}, \ref{vmu})  using
the AL1 interquark potential. Upper right panel:
$h_+(w)$ as before, $\xi_+(w)$ obtained from the values of $h_+(w)$ using
Eq.~(\ref{hiw}), $h_-(w)$ obtained from $\xi_+(w)$ using Eq.~(\ref{hiw}). We
also show the UKQCD lattice data by Bowler {\it et al.}~\cite{ukqcd02}. Lower
panel: Different $\xi_+(w)$ obtained with the above procedure for the different
interquark interactions. Lattice data is also shown.}
\label{fig:hpml}
\end{figure}

In the upper left panel of Fig.~\ref{fig:hpml} we show the values of
$h_+(w)$ and $h_-(w)$ for the $B\to D$
transition as obtained from Eqs.~(\ref{hpm}, \ref{vmu}) with the use of the AL1
interquark potential. The values for $h_-(w)$ 
are not reliable. Actual calculation shows that they are of the same size as 
the deviations from zero that
one computes in the equal mass case. 
To improve on this what we shall do instead is to use the
form factor $h_+(w)$ and Eq.~(\ref{hiw}) to extract $\xi(w)$ (we shall call it
$\xi_+(w)$\,) and from there we can
re-evaluate $h_-(w)$  with the use of Eq.~(\ref{hiw}). The results appear 
in the upper right panel of Fig.~\ref{fig:hpml} where we also show the lattice results
for $\xi(w)$ obtained by the UKQCD Collaboration in Ref.~\cite{ukqcd02}. 
We find good agreement with lattice data. Finally in the lower panel of Fig.~\ref{fig:hpml} 
we show the different $\xi_+(w)$ obtained with the use of the different interquark
potentials. As we see from the figure all $\xi_+(w)$ are very much the same in the
whole interval for $w$.

 The slope at the
origin of our Isgur-Wise function is given by
\begin{eqnarray}
\rho^2=-\frac{1}{\xi_+(w)}\left.\frac{d\,\xi_+(w)}{dw}\right|_{w=1}
=0.35 \pm 0.02
\end{eqnarray}
small  compared to the lattice value of $\rho^2=0.81^{+17}_{-11}$ extracted from
a best fit to data.

\subsubsection{Differential decay width}
Neglecting lepton masses the differential decay width for the process
$B\to D\,l\,\bar{\nu}$ is given by~\cite{neubert91}
\begin{eqnarray}
\frac{d\Gamma}{d w}=\frac{G_F^2}{48\pi^3}\,|V_{cb}|^2\,M_D^3\,
(w^2-1)^{3/2}(M_B+M_D)^2\ F_D^2(w)                                       
\end{eqnarray}
where $G_F=1.16637(1)\times
10^{-5}$\, GeV$^{-2}$~\cite{pdg04} is the Fermi decay constant, $V_{cb}$ is the
CKM matrix
element for the $b\to c$ weak transition, and $F_D(w)$ is given by
\begin{eqnarray}
F_D(w)=\bigg[h_+(w)-\frac{1-r}{1+r}\,h_-(w)\bigg] 
\end{eqnarray}
with $r=M_D/M_B$.

\begin{figure}[h!!!!]
\vspace{1cm}
\centering
\resizebox{11cm}{6cm}{\includegraphics{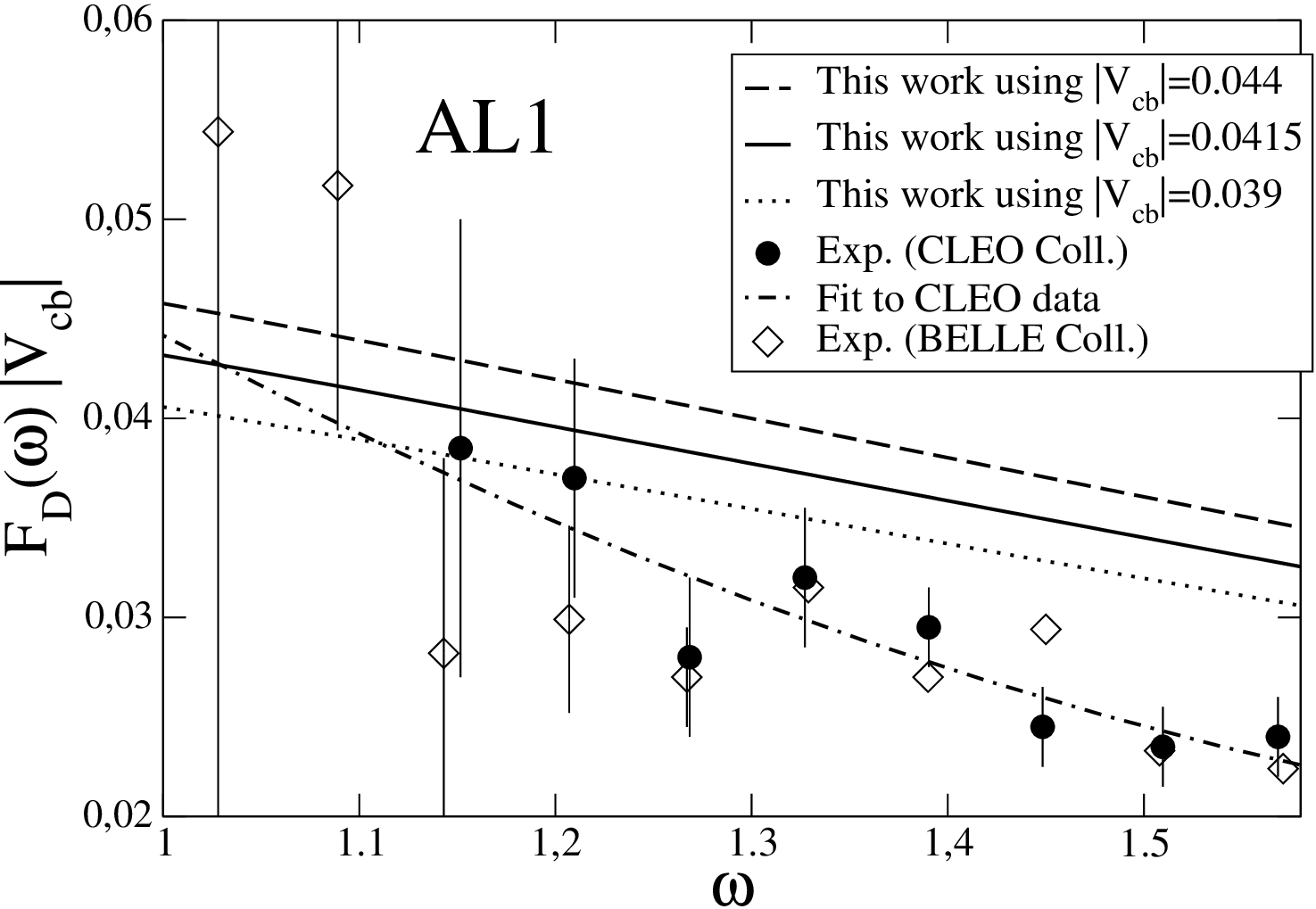}}
\caption{$F_D(r,w)\ |V_{cb}|$ obtained with the AL1 interquark potential.
 Solid line: our results
using $|V_{cb}|=0.0415$. Dashed line: our results
using $|V_{cb}|=0.044$. Dotted line: our results
using $|V_{cb}|=0.039$.
Circles: experimental data by the CLEO Collaboration~\cite{cleo99}.
Dashed-dotted line:
fit to CLEO data using the form factors of Boyd {\it et al.}~\cite{boyd97}.
Diamonds: experimental data by the BELLE Collaboration~\cite{belle02}. }
\label{fig:fd}
\end{figure}

In Fig.~\ref{fig:fd} we show our calculation for $F_D(w)\ |V_{cb}|$ obtained
with the AL1 interquark potential and using three
different values of  $|V_{cb}|$ corresponding to the 
central and extreme values of the range for $|V_{cb}|$ favored by the
Particle Data Group (PDG), 
 $|V_{cb}|=0.039 \sim 0.044$~\cite{pdg04}. We also show the experimental data 
 for the decays $B^-\to D^0\, l\,\bar{\nu}$ and 
$\bar{B}^0\to D^+\,l\,\bar{\nu}$ obtained by the CLEO
Collaboration~\cite{cleo99}, a fit to CLEO data using the form factors of
 Boyd {\it et al.}~\cite{boyd97}, and the experimental data for the decay 
 $\bar{B}^0\to D^+\,l\,\bar{\nu}$ obtained by the BELLE
Collaboration~\cite{belle02}. Our results are larger than experimental
data for $w >1.2$. Our total integrated width will thus be larger than
the experimental one for any reasonable value of $|V_{cb}|$. 
From our data we
extract the slope at $w=1$ given by
\begin{eqnarray}
\rho_D^2=\left.-\frac{1}{F_D(w)}\frac{d F_D(w)}{d w}\right|_{w=1}=
0.38 \pm 0.02
\end{eqnarray}
which is  small compared to the values extracted from the experimental data: 
 \hbox{$\rho_D^2=0.76\pm0.16\pm0.08$}~\cite{cleo99} and
\hbox{$\rho_D^2=0.69\pm0.14$}~\cite{belle02} obtained from a linear fit to the
data, or 
$\rho_D^2=1.30\pm0.27\pm0.14$~\cite{cleo99} and
$\rho_D^2=1.16\pm0.25$~\cite{belle02} obtained using the form factors of Boyd
{\it et al.}~\cite{boyd97}. Thus, only our
results close to $w=1$ seem to be reliable. We can use our prediction for
 $F_D(1)$ to extract the value of $|V_{cb}|$ from the experimental
determination of the quantity $|V_{cb}|F_D(1)$. Different values of that
quantity
appear in  Table~\ref{tab:vcbfd}
 \begin{table}[h!!!!]
\begin{center}
\begin{tabular}{l|c  }
 & $|V_{cb}|\,F_{D}(1)$ \\
\hline
& \\
CLEO Collaboration\hfill ~\protect{\cite{cleo99}} & $\ 0.0416\pm0.0047\pm0.0037$\\
BELLE Collaboration\hfill~\protect{\cite{belle02}} & $\ 0.0411\pm0.0044\pm0.0052$
\end{tabular}
\end{center}
\caption{ $|V_{cb}|\,F_{D}(1)$ values obtained by different experiments. }
\label{tab:vcbfd}
\end{table}

Our result for  $F_D(1)$ is given by (we do not show the theoretical
 error which
is or the order of $10^{-4}$)
\begin{eqnarray}
F_D(1)=1.04
\end{eqnarray}
which is in good agreement with other calculations
$F_D(1)=0.98\pm0.07$~\cite{caprini96}, $F_D(1)=1.04$~\cite{scora95} or
\hbox{$F_D(1)=1.069\pm0.008\pm0.002\pm0.025$~\cite{hashimoto99}}.
From our value for $F_D(1)$ and the experimental values for $|V_{cb}|F_D(1)$
we can obtain  $|V_{cb}|$  in the range
\begin{eqnarray}
|V_{cb}|=0.040\pm0.006
\end{eqnarray}
This result agrees with our recent determination based on the analysis of the 
$\Lambda_b\to\Lambda_c\,l\,\bar{\nu}_l$ reaction from where we got 
$|V_{cb}|=0.040\pm0.005$~\cite{albertus05}.

\subsection{$B\to D^*\,l\,\bar{\nu}$ decay}
\label{subsect:btod*}
Working again in the center of mass of  the $B$ meson and taking $\vec{P}^{\,\prime}=
-\vec{q}=-|\vec{q}\,|\,
\widehat{\vec{k}}$ in the z 
direction  we will have for the form factors $h_V(w)$, $h_{A_1}(w)$, 
$h_{A_2}(w)$ and $h_{A_3}(w)$ the expressions
\begin{eqnarray}
\label{hva123}
h_V(w)\hspace{0.0cm}&=&\sqrt2\,\sqrt{\frac{M_{D^*}}{M_B}}\,\frac{V^{(*)}_{-1,\,2}(|\vec{q}\,|)}{|\vec{q}\,|}\nonumber\\
h_{A_1}(w)&=&i\,\frac{\sqrt2}{w+1}\,\frac{1}{\sqrt{M_BM_{D^*}}}\,
A^{(*)}_{-1,\, 1}(|\vec{q}\,|)\nonumber\\
h_{A_2}(w)&=&i\,\sqrt{\frac{M_{D^*}}{M_B}}\,
\left(-\frac{A^{(*)}_{0,\, 0}(|\vec{q}\,|)}{|\vec{q}\,|}+
\frac{E_{D^*}(|\vec{q}\,|)\,A^{(*)}_{0,\, 3}(|\vec{q}\,|)}{|\vec{q}\,|^2}
-\sqrt2\,M_{D^*}
\frac{A^{(*)}_{-1,\, 1}(|\vec{q}\,|)}{|\vec{q}\,|^2}
\right)\nonumber\\
h_{A_3}(w)&=&i\frac{M_{D^*}^2}{\sqrt{M_BM_{D^*}}}
\left(-\frac{A^{(*)}_{0,\, 3}(|\vec{q}\,|)}{|\vec{q}\,|^2}+\,
\frac{\sqrt2}{M_{D^*}}
\frac{E_{D^*}(|\vec{q}\,|)\,A^{(*)}_{-1,\, 1}(|\vec{q}\,|)}{|\vec{q}\,|^2}\right)
\end{eqnarray}
with $E_{D^*}(|\vec{q}\,|)=\sqrt{M_{D^*}^2+|\vec{q}\,|^2}$, and
where $V^{(*)}_{\lambda,\,\mu}(|\vec{q}\,|)$ and 
$A^{(*)}_{\lambda,\,\mu}(|\vec{q}\,|)$
are given by
\begin{eqnarray}
V^{(*)}_{\lambda,\,\mu}(|\vec{q}\,|)=\left\langle\, D^*,\,\lambda\ -|\vec{q}\,
|\,\vec{k}\,\left|\, (J^{c\,b}_V)_\mu(0)\,
\right| \, B,\,\vec{0}\right\rangle\nonumber\\
A^{(*)}_{\lambda,\,\mu}(|\vec{q}\,|)=\left\langle\, D^*,\,\lambda\ -|\vec{q}\,
|\,\vec{k}\,\left|\, (J^{c\,b}_A)_\mu(0)\,
\right| \, B,\,\vec{0}\right\rangle
\end{eqnarray}
In our model $V^{(*)}_{\lambda,\,\mu}(|\vec{q}\,|)$ and 
$A^{(*)}_{\lambda,\,\mu}(|\vec{q}\,|)$ are evaluated as
\begin{eqnarray}
\label{va123}
V^{(*)}_{\lambda,\,\mu}(|\vec{q}\,|)&=&\sqrt{2M_B2E_{D^*}(|\vec{q}\,|)}
\ \ {}_{\stackrel{}{NR}}
\left\langle\, D^*,\,\lambda\ 
-|\vec{q}\,|\,\widehat{\vec{k}}\,\left|\, (J^{c\,b}_V)_\mu(0)\,
\right| \, B,\,\vec{0}\right\rangle_{NR}\nonumber\\
A^{(*)}_{\lambda,\,\mu}(|\vec{q}\,|)&=&\sqrt{2M_B2E_{D^*}(|\vec{q}\,|)}\ \ 
{}_{\stackrel{}{NR}}
\left\langle\, D^*,\,\lambda\ 
-|\vec{q}\,|\,\widehat{\vec{k}}\,\left|\, (J^{c\,b}_A)_\mu(0)\,
\right| \, B,\,\vec{0}\right\rangle_{NR}\nonumber\\
\end{eqnarray}
with expressions given in appendix~\ref{app:va}.

In the left panel of Fig.~\ref{fig:hva123r12} we show our results for the $h_V(w), h_{A_1}(w), 
h_{A_2}(w)$ and $h_{A_3}(w)$ form factors, obtained with the AL1 interquark
potential and the use of Eq.~(\ref{hva123}). In the right panel of the same figure we
show the ratios
\begin{figure}[t]
\vspace{1cm}
\centering
\resizebox{7cm}{5cm}{\includegraphics{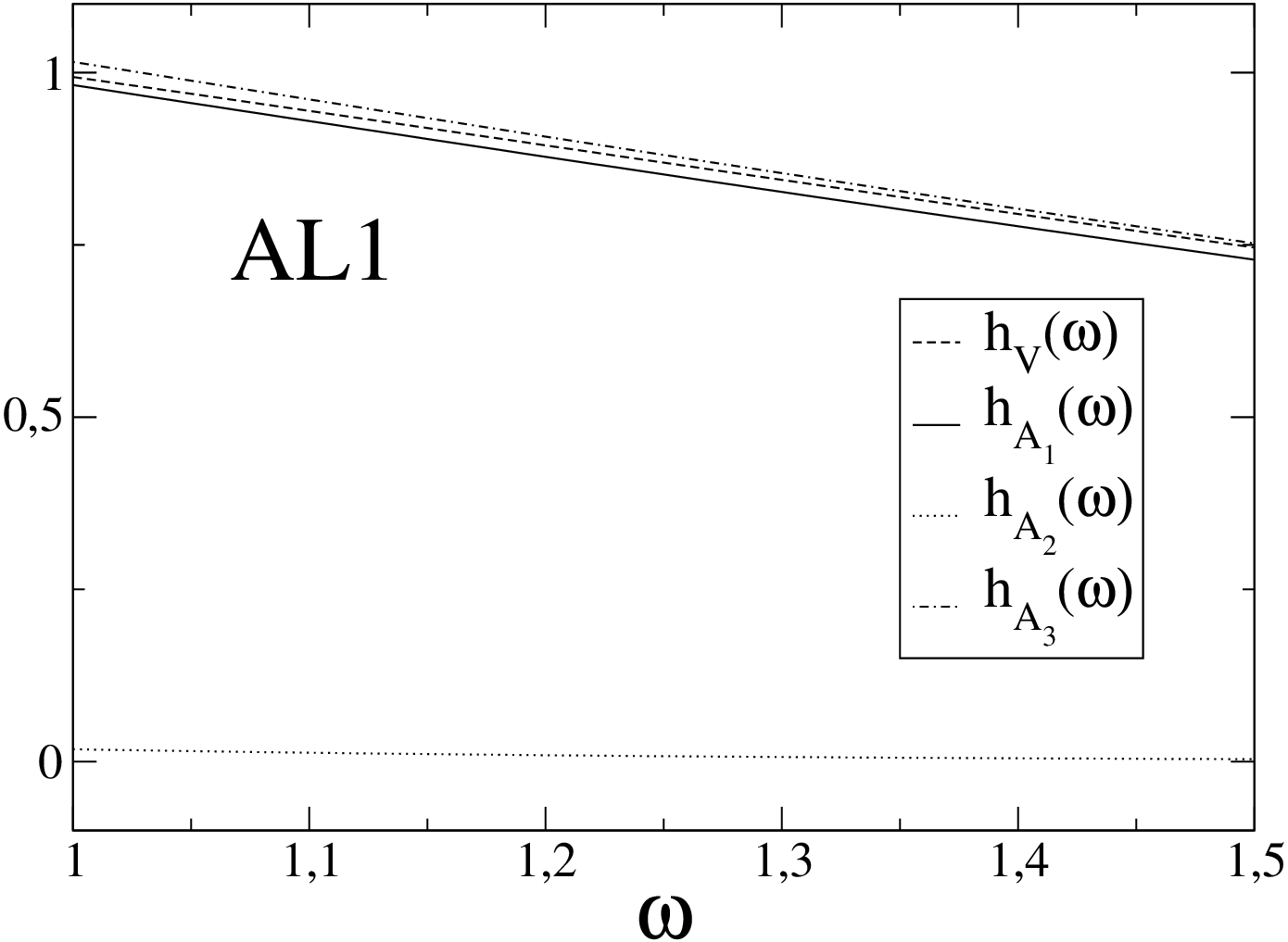}}\hspace{2cm}
\resizebox{7cm}{5cm}{\includegraphics{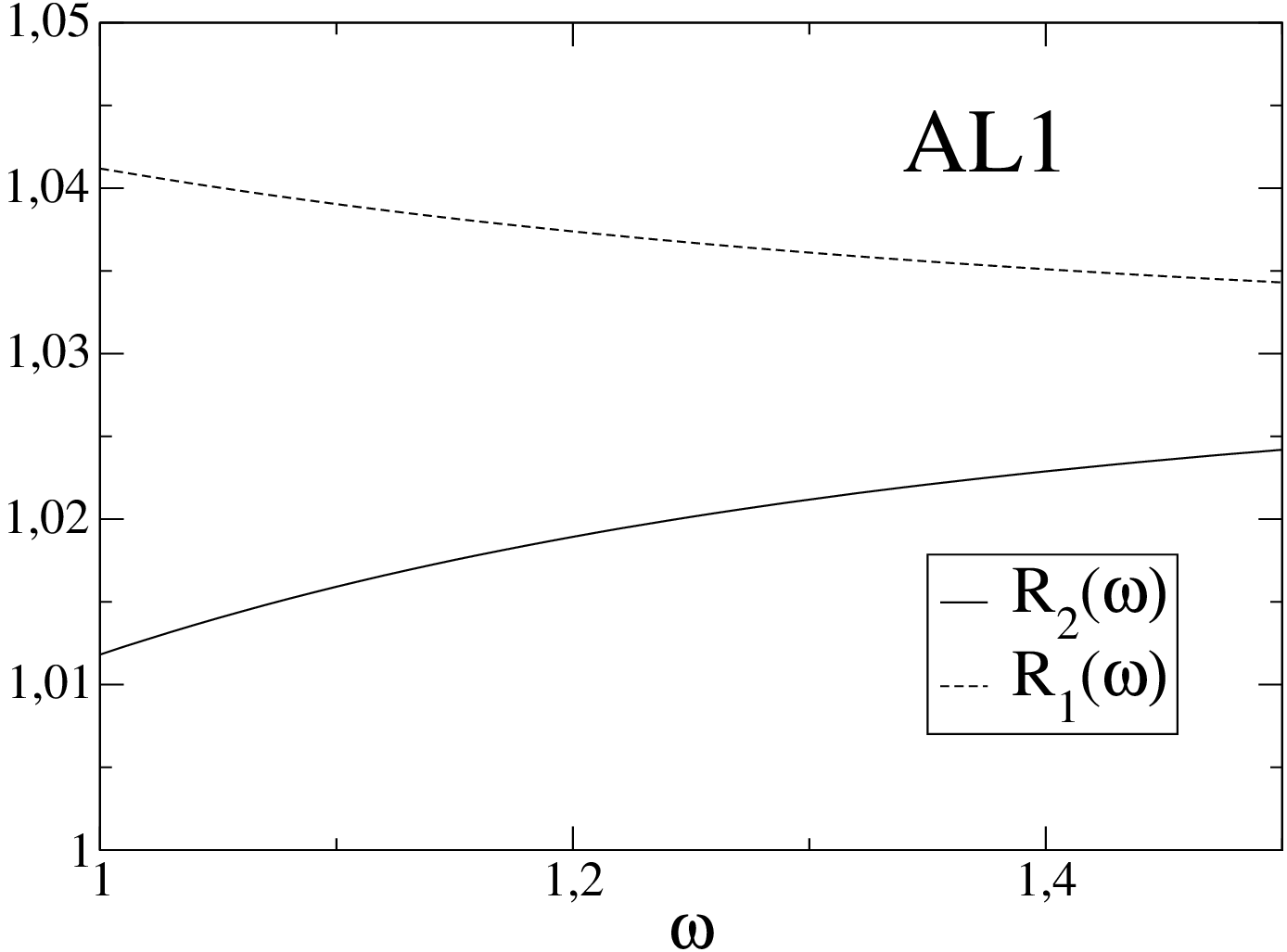}}
\caption{Left panel: $h_V(w)$, $h_{A_1}(w)$, $h_{A_2}(w)$ and $h_{A_3}(w)$ form
factors obtained  using Eq.~(\ref{hva123}). Right panel: 
$R_1(w)$ and $R_2(w)$ ratios. In both panels the AL1 interquark potential has
been used. }
\label{fig:hva123r12}
\end{figure}
\begin{eqnarray}
&&R_1(w)=\frac{h_V(w)}{h_{A_1}(w)}\nonumber\\
&&R_2(w)=\frac{h_{A_3}(w)+r\,h_{A_2}(w)}{h_{A_1}(w)}
\end{eqnarray}
where now $r=M_{D^*}/M_B$. These ratios are expected to vary very weakly with
$w$.
We find indeed that this  is so in our case being our values of $R_1(w)$ 
and $R_2(w)$ 
within 4\% of  unity.
In Table~\ref{tab:r12} we give now our results for $R_1(1)$ and $R_2(1)$ and compare
them to different experimental\footnote{The experimental
results by the CLEO and BABAR Collaborations have been  obtained with the
assumption that $R_1(w)$ and $R_2(w)$ are constants.} and theoretical 
determinations. We find discrepancies of
the order of $15\sim 33\, \% 
$ 
for $R_1(1)$ and $13\sim 46\, \%
 $ for $R_2(1)$. 
 
\begin{figure}[h!!!]
\vspace{.8cm}
\centering
\resizebox{7cm}{5cm}{\includegraphics{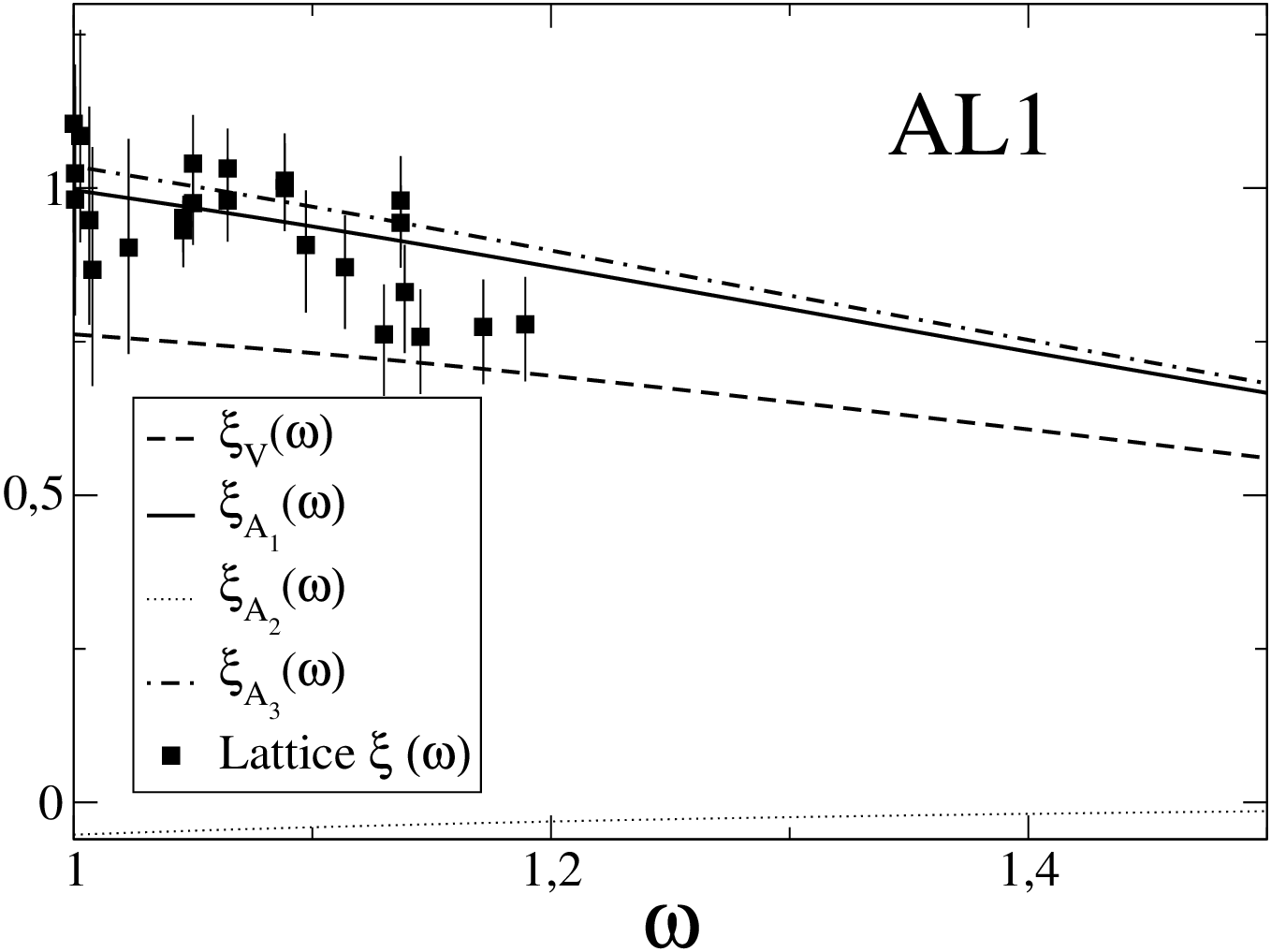}}\hspace{1cm}
\resizebox{7cm}{5cm}{\includegraphics{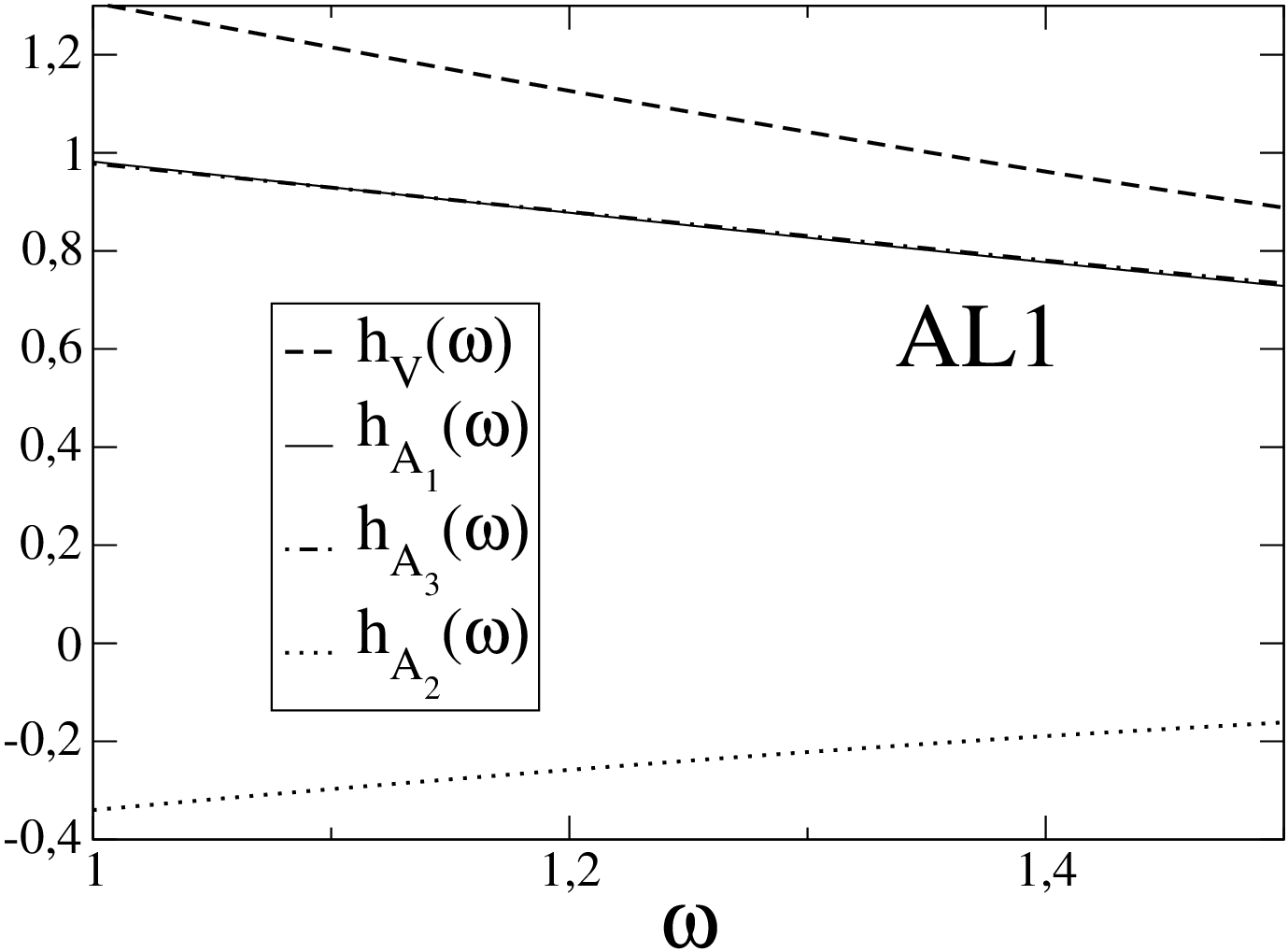}}\vspace{.7cm}\\
\resizebox{7cm}{5cm}{\includegraphics{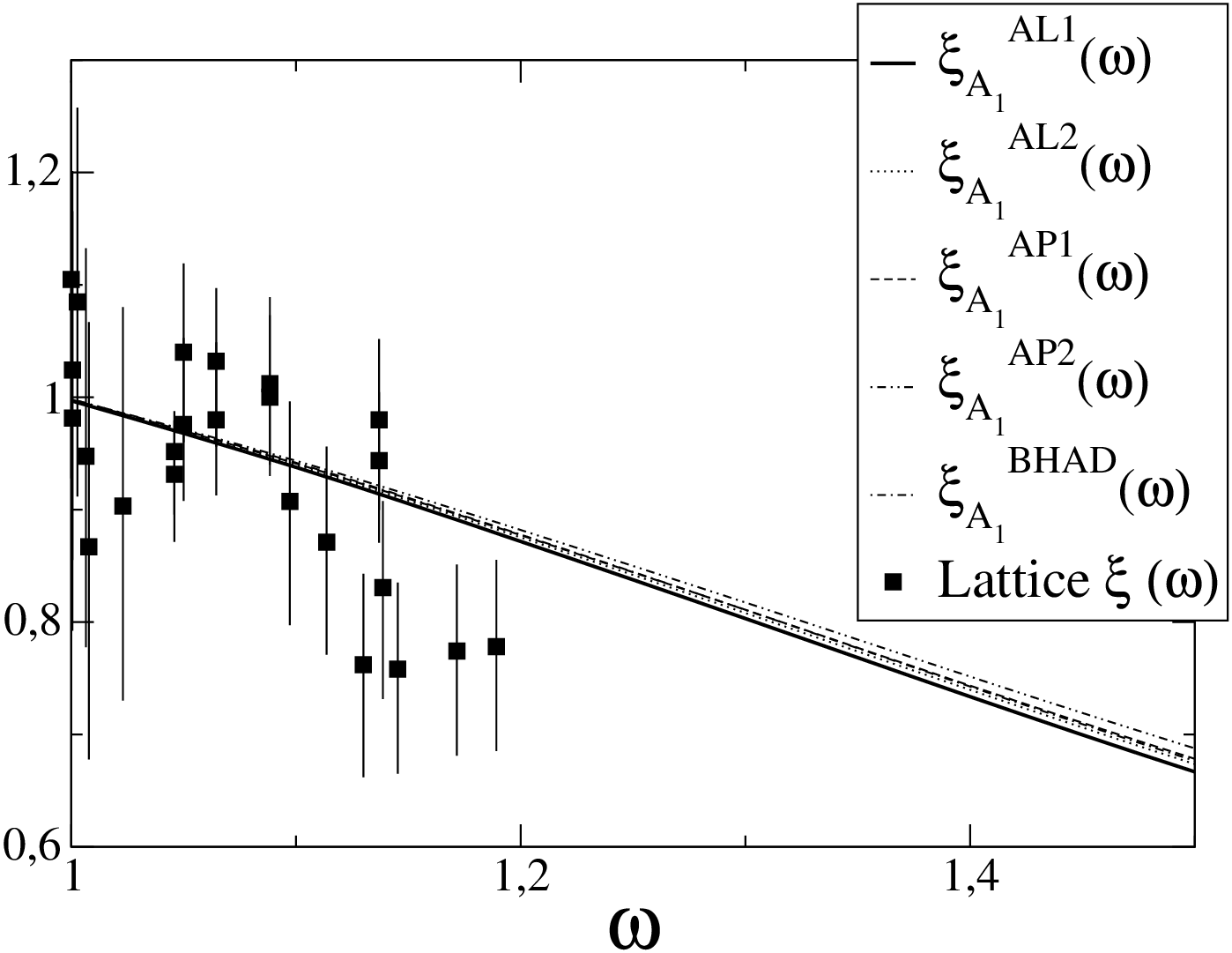}}
\caption{Upper left panel: different $\xi(w)$ functions obtained from the $h_V(w)$, 
$h_{A_1}(w)$, $h_{A_2}(w)$ and $h_{A_3}(w)$ form
factors
using Eq.~(\ref{hiw}) and the AL1 interquark potential. Lattice data  by K. C. Bowler {\it et
al.} from
Ref.~\cite{ukqcd02} are also shown. Upper right panel: form factors obtained  from
$\xi_{A_1}(w)$ with the use of Eq.~(\ref{hiw}). Lower panel: Different 
$\xi_{A_1}(w)$ obtained with the different interquark potentials.}
\label{fig:xiva123}
\end{figure}

One can understand these discrepancies by evaluating the different $\xi(w)$
functions obtained from the  form factors with the use of 
Eq.~(\ref{hiw})
and the $\beta$ and $\gamma$ coefficients of Neubert given in
 Tables~\ref{tab:beta} and \ref{tab:gamma}.
The results appear in the upper left panel of Fig.~\ref{fig:xiva123}. One can 
infer from the
figure that our results for $h_{A_2}(w)$ are not reliable. 
Also we somehow miss the
correct normalization for $h_{V}(1)$. On the other hand the values of 
$\xi_ {A_1}(w)$ and $\xi_{A_3}(w)$ are equal within 4\% and in reasonable
agreement with   lattice data from Ref.~\cite{ukqcd02}. 

To improve the nonrelatistic quark model prediction,
 and similarly to what we did in subsection~\ref{subsect:btod},
we will take $\xi_{A_1}(w)$ as our model determination of the
Isgur-Wise function $\xi(w)$ and we will reevaluate the form factors with the
use of Eq.~(\ref{hiw}). What we obtain is now depicted in the upper right panel of
Fig.~\ref{fig:xiva123}. In the lower panel we give the different $\xi_{A_1}(w)$ obtained
with the different interquark potentials. They do not show any significant
difference.

The slope of the $\xi_{A_1}(w)$ function at the origin is  given by
\begin{eqnarray}
\rho^2=0.55 \pm 0.02
\end{eqnarray}
to be compared to the lattice result $\rho^2=0.93^{+47}_{-59}$~\cite{ukqcd02}.
In this case we are within lattice errors, but one
can not be very conclusive due to the 
large value of the latter in this case.
\begin{table}[t]
\begin{center}
\begin{tabular}{c|c c }
 & $R_1(1)$  & $R_2(1)$\\
\hline
This work &  $1.01\pm0.02$ & $1.04\pm0.01$\\
CLEO~\cite{cleo03} & $1.18\pm0.30\pm0.12$ & $0.71\pm0.22\pm0.07$ \\
BABAR (Preliminary)~\cite{babar04} &\hspace{.5cm}$1.328\pm0.055\pm0.025\pm0.025$\hspace{.5cm} &
$0.920\pm0.044\pm0.020\pm0.013$ \\ \hline
& &\\
Caprini {\it et al.}~\cite{caprini98}& 1.27 & 0.80\\
Grinstein {\it et al.}~\cite{grinstein02}& 1.25& 0.81\\
Close {\it et al.}~\cite{close94}& 1.15& 0.91
\end{tabular}
\end{center}
\caption{ $R_1(1)$ and $R_2(1)$ }
\label{tab:r12}
\end{table}

\begin{figure}[h!!!]
\vspace{.8cm}
\centering
\resizebox{9cm}{7cm}{\includegraphics{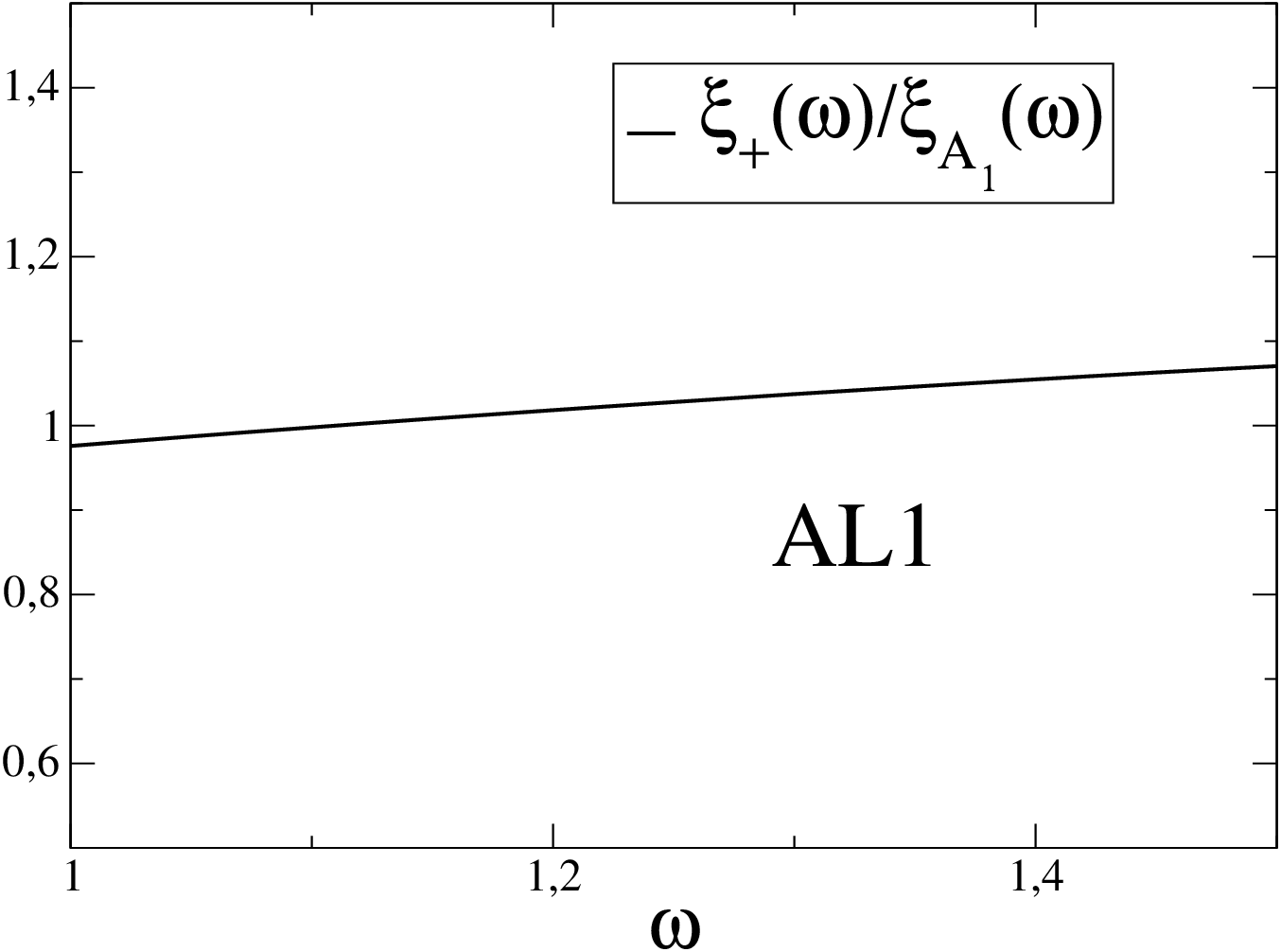}}
\caption{Ratio of our two Isgur-Wise functions calculated with the AL1
interquark potential.}
\label{fig:xipxia1}
\end{figure}

Finally in Fig.~\ref{fig:xipxia1} we give the ratio $\xi_+(w)/\xi_{A_1}(w)$
evaluated with the AL1 interquark potential. We
see  the differences between the two Isgur-Wise functions are at the level of 
3-7\%.

\subsubsection{Differential decay width}
Neglecting lepton masses the differential decay width for the process
$B\to D^*\,l\,\bar{\nu}$ is given by~\cite{richman95}
\begin{eqnarray}
\frac{d\Gamma}{d w}=\frac{G_F^2}{48\pi^3}\,|V_{cb}|^2\,(M_B-M_{D^*})^2\,M_{D^*}^3\,
\sqrt{(w^2-1)}\,(w+1)^2\bigg[
1+\frac{4w}{w+1}\ \frac{1-2wr+r^2}{(1-r)^2}
\bigg]\ F_{D^*}^2(w)                                       
\end{eqnarray}
where  $F_{D^*}(w)$ is defined as
\begin{eqnarray}
F_{D^*}(w)=h_{A_1}(w)\sqrt{\frac{\tilde{H}_0^2(w)+\tilde{H}_+^2(w)+
\tilde{H}_-^2(w)}
{1+\frac{4w}{w+1}\ \frac{1-2wr+r^2}{(1-r)^2}}} 
\end{eqnarray}
The $\tilde{H}_j(w)$ are helicity form factors given in terms of the $R_1(w)$ 
and $R_2(w)$ ratios as
\begin{eqnarray}
\tilde{H}_0(w)&=&1+\frac{w-1}{1-r}\,[1-R_2(w)]\nonumber\\
\tilde{H}_\pm(w)&=&\frac{\sqrt{1-2wr+r^2}}{1-r}\left[
1\mp\sqrt{\frac{w-1}{w+1}}\ R_1(w)
\right]
\end{eqnarray}

\begin{figure}[h!!!]
\vspace{1cm}
\centering
\resizebox{13cm}{7cm}{\includegraphics{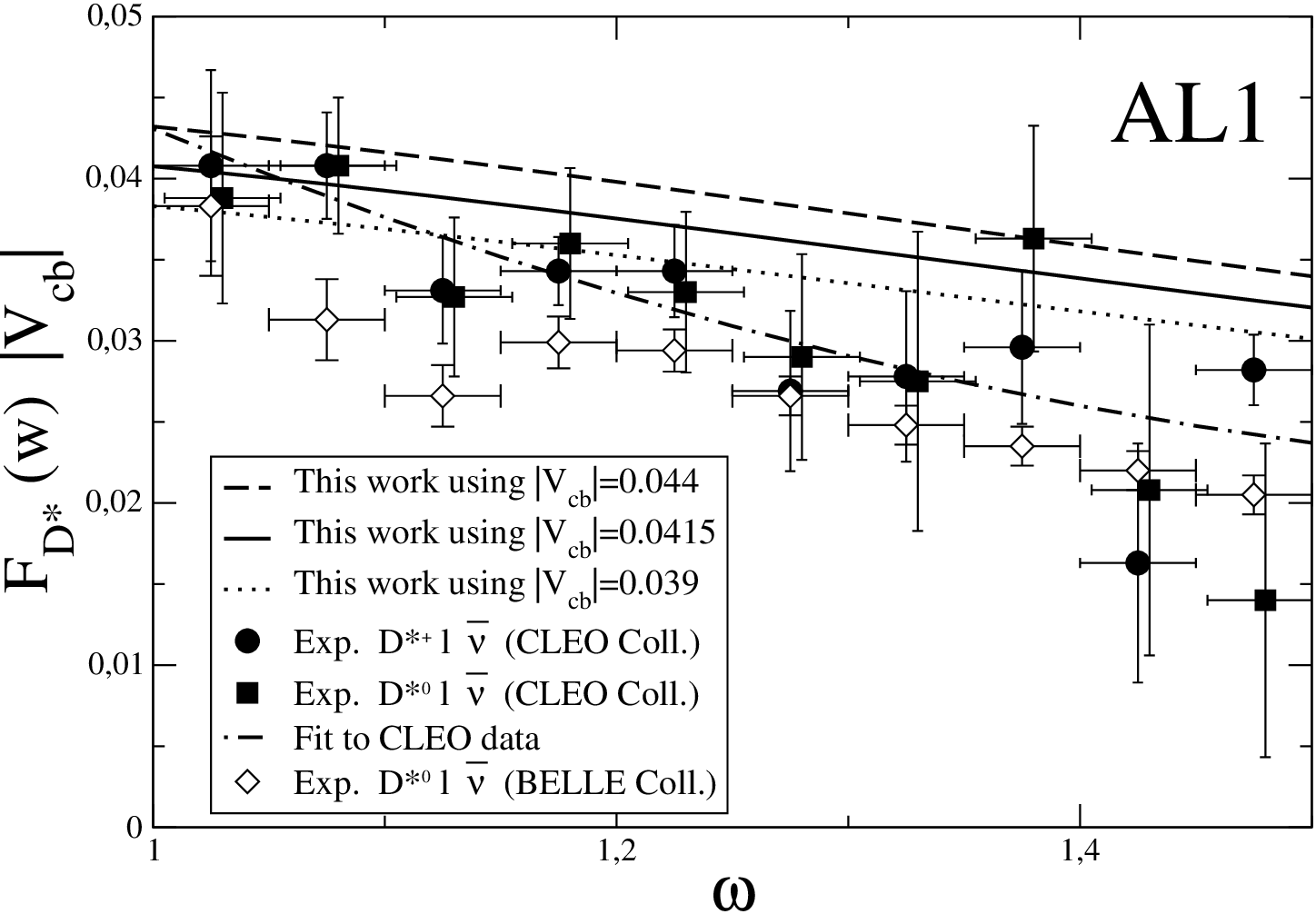}}
\caption{$F_{D^*}(r,w)\ |V_{cb}|$ obtained with the AL1 interquark potential.
 Solid line: our results
using $|V_{cb}|=0.0415$. Dashed line: our results
using $|V_{cb}|=0.044$. Dotted line: our results
using $|V_{cb}|=0.039$.
Circles and squares: experimental data by the CLEO Collaboration~\cite{cleo03}.
Dashed-dotted line: fit to CLEO Collaboration data. Diamonds: experimental data by the
BELLE Collaboration~\cite{belle02s}.}
\label{fig:fdstar}
\end{figure}

Similarly to Fig.~\ref{fig:fd}, in Fig.~\ref{fig:fdstar} we show our results 
for the quantity  $F_{D^*}(w)\
|V_{cb}|$ evaluated with the AL1 interquark potential and using the values of $|V_{cb}|$ corresponding to the central and
extreme values of the range for  $|V_{cb}|$  favored by the PDG. We also show the experimental data by the CLEO
 Collaboration~\cite{cleo03} for the $B^-\to D^{*0}\,l\,\bar{\nu}$ reaction
  (squares),
  and for the  $\bar{B}^0\to D^{*+}\,l\,\bar{\nu}$ reaction
 (circles) together with a best fit, and the
 experimental data by the BELLE Collaboration~\cite{belle02s} for the $\bar{B}^0\to D^{*+}\,l\,\bar{\nu}$ 
 reaction (diamonds). We find 
 good agreement
 with CLEO data for small $w$ values. Disagreement starts already
 at around $w=1.1$ where our resuls start to go above the experimental data.
   BELLE data are systematically below our results.
  
 Also our slope at the origin
 \begin{eqnarray}
 \rho^2_{D^*}=-\frac{1}{F_{D^*}(w)}\left.\frac{d F_{D^*}(w)}{d w}\right|_{w=1}
 =0.31 \pm 0.02
 \end{eqnarray}
 is  smaller  than the value
  obtained by the BELLE Collaboration
  \hbox{$\rho^2_{D^*}=0.81\pm0.12$}~\cite{belle02s} using a linear fit to their
  data.
 All this means that our total width would be larger than the 
 experimental one for any reasonable value of $V_{cb}$. On the other hand
  experimentalists are able to extract the value of $|V_{cb}|\,F_{D^*}(1)$.
 Different experimental results for that quantity appear in
 Table~\ref{tab:vcbfds}. 
 \begin{table}[h!!!!]
\begin{center}
\begin{tabular}{l|c  }
 & $|V_{cb}|\,F_{D^*}(1)$ \\
\hline
& \\
(CLEO Coll.)\ \hfill \protect{\cite{cleo03}} & $\ 0.0431\pm0.0013\pm0.0018$\\
(DELPHI Coll.)\ \hfill\protect{\cite{delphi04}} & $\ 0.0392\pm0.0018\pm0.0023$\\
(BELLE Coll.)\ \hfill\protect{\cite{belle02s}}& $\ 0.0354\pm0.0019\pm0.0018$\\
(BABAR Coll)\ \hfill\protect{\cite{babar05}} & $\ 0.0355\pm0.0003\pm0.0016$ 
\end{tabular}
\end{center}
\caption{ $|V_{cb}|\,F_{D^*}(1)$ values obtained by different experiments. }
\label{tab:vcbfds}
\end{table}

Our result  for  $F_{D^*}(1)$ is given by 
  \begin{eqnarray}
 F_{D^*}(1)=h_{A_1}(1)=0.983\pm 0.001
 \end{eqnarray}
 Comparison with the experimental data for $|V_{cb}|\,F_{D^*}(1)$ allows us to 
  extract
  values for $|V_{cb}|$ in the range
 \begin{equation}
 |V_{cb}|=0.0333\sim0.0461
 \end{equation}
 One can not be more conclusive due to  the dispersion in the experimental data
 for $|V_{cb}|\,F_{D^*}(1)$.  From DELPHI data alone we would obtain 
 $|V_{cb}|=0.040\pm0.003$ in perfect agreement with our determination using
 the $B\to D$ reaction data.  We
 should also say that our value for $F_{D^*}(1)$
is larger than the lattice determination 
$F_{D^*}(1)=0.919^{+0.030}_{-0.035}$ by S. Hashimoto {\it et al.}~\cite{hashimoto02}
normally  used
by experimentalists to extract their $|V_{cb}|$ values. A new unquenched lattice
determination of this quantity by the Fermilab Lattice
Collaboration is in progress~\cite{hashimoto04}.

\section{Strong coupling constants $g_{H^*H\pi}$}
\label{sect:gh*hpi}
In this section we will evaluate the strong coupling constants $g_{H^*H\pi}$ 
where $H$ stands for a $B$ or $D$ meson. To this end we shall make use of the
partial conservation of the axial current hypothesis (PCAC) which relates the
divergence of the axial current to the pion field as
\begin{eqnarray}
\label{pcac}
\partial^\mu\, J_{A\,\mu}^{d\, u}(x)=\,f_{\pi}\,m_{\pi}^2\, \Phi_{\pi^-}(x)
\end{eqnarray}
where $f_{\pi}=130.7\pm0.1\pm0.36 $\,MeV~\cite{pdg04} is the pion decay constant, $m_{\pi}=
139.57$\,MeV~\cite{pdg04} is the pion mass, and $\Phi_{\pi^-}(x)$ is the charged pion
field that destroys a $\pi^-$ and creates a $\pi^+$. Using the LSZ reduction 
formula one can  relate the matrix element
of the divergence of the axial current to the pion emission amplitude as
\begin{eqnarray}
\label{pcac2}
\langle H,\ \vec{P}^\prime\,|\,q^\mu\, J_{A\,\mu}^{d\, u}(0)\,|
\,H^*,\lambda\,\vec{P}\,\rangle= -i\,f_{\pi}\frac{m_{\pi}^2}{q^2-m_{\pi}^2}\ 
{\cal A}^{(\lambda)}_{H^*H\pi}(P^\prime,P)
\end{eqnarray}
where $q=P-P^\prime$ and ${\cal A}^{(\lambda)}_{H^*H\pi}(P^\prime,P)$ is 
the pion emission amplitude for the process $H^*\to H\pi$  given
by\footnote{Corresponding to the emission  of a $\pi^+$}
\begin{eqnarray}
{\cal A}^{(\lambda)}_{H^*H\pi}(P^\prime,P)=-\,g_{H^*H\pi}(q^2)\,
\left(q^{\mu}\,\varepsilon^{(\lambda)}_{\mu}(\vec{P}\,)\right)
\end{eqnarray}
The matrix element on the left hand side of Eq.(\ref{pcac2}) has a pion pole
contribution that can be easily evaluated to be
\begin{eqnarray}
\label{pp}
\langle H,\ \vec{P}^\prime\,|\,q^\mu\, J_{A\,\mu}^{d\, u}(0)\,|
\,H^*,\lambda\,\vec{P}\,\rangle_{pion-pole}= -i\,f_{\pi}\frac{q^2}{q^2-m_{\pi}^2}\ 
{\cal A}^{(\lambda)}_{H^*H\pi}(P^\prime,P)
\end{eqnarray}
so that we can extract a non-pole contribution
\begin{eqnarray}
\label{np}
\langle H,\ \vec{P}^\prime\,|\,q^\mu\, J_{A\,\mu}^{d\, u}(0)\,|
\,H^*,\lambda\,\vec{P}\,\rangle_{non-pole}&=& i\,f_{\pi}\ 
{\cal A}^{(\lambda)}_{H^*H\pi}(P^\prime,P)\nonumber\\
&=&-i\,f_{\pi}\ g_{H^*H\pi}(q^2)
\,\left(q^{\mu}\,\varepsilon^{(\lambda)}_{\mu}(\vec{P}\,)\right)
\end{eqnarray}
which is the one we shall evaluate within the quark model.
For the matrix element on the left hand side of Eq.(\ref{np}) we can use a
parametrization similar to the one used in Eq.({\ref{bd*}})
\begin{eqnarray}
\langle H,\ \vec{P}^\prime\,|\,q^\mu\, J_{A\,\mu}^{d\, u}(0)\,|
\,H^*,\lambda\,\vec{P}\,\rangle_{non-pole}=q^\mu\bigg\{\hspace{-.4cm}&&-i\,
\varepsilon^{(\lambda)}_{\mu}(\vec{P}\,)\ (w+1)\ h_{A_1}(w)\nonumber\\
&&+i \left(\varepsilon^{(\lambda)}(\vec{P}\,)\cdot v^\prime\right)
\left( v^\prime_\mu\, h_{A_2}(w)+v_\mu\,h_{A_3}(w) 
\right)\bigg\} \sqrt{M_H M_{H^*}}
\end{eqnarray}
with the result that
\begin{eqnarray}
g_{H^*H\pi}(q^2)=\frac{1}{f_{\pi}}\bigg\{
(w+1)\,h_{A_1}(w)+w\left(\frac{M_{H^*}}{M_H}\,h_{A_2}(w)-h_{A_3}(w)\right)\nonumber\\
+\left(\frac{M_{H^*}}{M_H}\,h_{A_3}(w)-h_{A_2}(w)\right)
\bigg\} \sqrt{M_H M_{H^*}}
\end{eqnarray}
The evaluation of the form factors is done is a similar way as the one
described in subsection~\ref{subsect:btod*}. The results that we get for
$q^2=0$ are
\begin{eqnarray}
g_{D^*D\pi}(0)=22.1\pm 0.4\hspace{.5cm},\hspace{.5cm}g_{B^*B\pi}(0)=60.5\pm1.1
\end{eqnarray}
to be compared to the experimental determination 
$g_{D^*D\pi}(m_\pi^2)=17.9\pm0.3\pm1.9$ by the CLEO Collaboration~\cite{cleo02},
the lattice results
$g_{D^*D\pi}(m_\pi^2)=18.8\pm2.3^{+1.1}_{-2.0}$~\cite{abada02}
 and $g_{B^*B\pi}(0)=47\pm5\pm8$~\cite{abada04}, or  a recent
 determination using QCDSR for which
 $g_{D^*D\pi}(m_\pi^2)=14.0\pm1.5$ and
 $g_{B^*B\pi}(0)=42.5\pm2.6$~\cite{navarra02}. Older QCDSR results give 
 smaller values for both coupling
 constants. For instance, the calculation within QCDSR on the light cone in 
 Ref.~\cite{belyaev95} give
  $g_{D^*D\pi}(m_\pi^2)=12.5\pm1$ and $g_{B^*B\pi}(0)=29\pm3$\ \footnote{
  Values for both coupling
  constants obtained prior to 1995 within different approaches  can be found
  in Ref.~\cite{belyaev95} and references therein }. The latter are
   small compared to lattice data or the experimental determination of
  $g_{D^*D\pi}(m_\pi^2)$ by the CLEO Collaboration.
 
From our results we obtain the ratio
\begin{eqnarray}
R=\frac{g_{B^*B\pi}(0)\ f_{B^*}\,\sqrt{M_D}}
{g_{D^*D\pi}(0)\ f_{D^*}\,\sqrt{M_B}}=1.105\pm 0.005
\label{R}
\end{eqnarray}
in good agreement with  HQS that predicts a value of 1
with
 $1/m_Q$ corrections appearing  in next to leading
 order~\cite{burdman94}
 \footnote{Note  the  strong coupling constant used in 
 Ref.~\cite{burdman94}
 is  given in terms of ours as
 $(g_{H^*H\pi}\,f_{\pi})/(2 M_{H^*})$ with $H=B,D$}. Our result in Eq.(\ref{R}) is also in agreement with the one
 obtained combining lattice data for $f_{B^*}$ and $f_{D^*}$ from 
 Ref.~\cite{ukqcd01}, for $g_{B^*B\pi}(0)$ from Ref.~\cite{abada02}, and the
 experimental CLEO
 Collaboration data for $g_{D^*D\pi}(m_{\pi}^2)$ from Ref.~\cite{cleo02}. In
 this case one gets 
 \begin{equation}
 \left.R\right|_{Exp.-Latt.}=1.26\pm0.36
 \end{equation}
  where we have added errors in quadratures. A calculation using
 light cone QCDSR gives~\cite{belyaev95}
 \begin{equation}
 \left.R\right|_{QCDSR}=0.92
 \end{equation}

\section{Concluding remarks}
\label{sect:conclusions}
Our analysis of leptonic decay constants of mesons with a heavy 
$c$ or $b$ quark shows that in a nonrelativistic calculation the equality
$f_VM_V= f_PM_P$
is satisfied within 2\%. This equality is expected in HQS in the limit where the
heavy quark masses go to infinity. The nonrelativistic result  suggests that
the HQS infinite mass limit
sets in already at the $m_c$ scale, something that 
is not supported by lattice data for $D$ mesons where one finds
deviations as large as 20\% from the above equality.

One also finds problems in  the semileptonic $B\to D$ and $B\to D^*$ decays. We have seen 
how the $h_-(w)$ form factor of  the $B\to D$ decay and the
$h_{A_2}(w)$ form factor of the $B\to D^*$ decay are not reliably calculated in the
nonrelativistic quark model where one finds large deviations from the HQET
relations of Eq.(\ref{hiw}). We have tried to remedy this failure by evaluating the
Isgur-Wise function from a form factor whose calculation we trusted in the quark
model, $h_+(w)$ for the $B\to D$ decay, and $h_{A_1}(w)$ for the $B\to D^*$ 
decay. Those Isgur-Wise function were later used  together with   HQET 
constraints in Eq.(\ref{hiw}) to recalculate all other form factors. The two Isgur-Wise functions thus
determined show an overall reasonable agreement with lattice data but in both
cases the slope  seems to be too small. This deficiency multiply
its effects when one goes to larger $w$ values and as a consequence the quantities 
$F_D(w)\,|V_{cb}|$ and
$F_{D^*}(w)\,|V_{cb}|$  go above experimental data for $w$ values larger than 
1.2, and for any reasonable value of $|V_{cb}|$. A failure of some kind is  
 expected in a
nonrelativistic calculation as $w$ increases. For $w=1.2$ the three momenta of the final meson
amounts to 66\% ot its mass and relativistic corrections in the wave function
could start to be important. On the other hand we believe our results are sound at
zero recoil ($w=1$). That enables us to  obtain
$|V_{cb}|=0.040\pm0.006$ from the $B\to D$ decay, in perfect agreement with the
value $|V_{cb}|=0.040\pm0.005$ previously obtained by us from the analysis of 
the $\Lambda_b\to
\Lambda_c$ semileptonic decay,
 and in good agreement with a recent determination $|V_{cb}|=0.0414\pm0.0012\pm0.0021\pm0.0018$ by the
DELPHI Collaboration. The
experimental situation concerning the $B\to D^*$ reaction is not so clear as
different experiments give values for $F_{D^*}(w)\,|V_{cb}|$ which are hardly
compatible. From DELPHI Collaboration data we would get 
$|V_{cb}|=0.040\pm0.003$
in agreement with the above result.

Finally we have made use of  PCAC to evaluate the strong coupling constants
$g_{B^*B\pi}$ and $g_{D^*D\pi}$. Our results are larger than experimental data 
or the results provided by lattice and QCDSR calculations. In this case 
 the final meson is nearly at rest and one would  expect a nonrelativistic 
 calculation to perform better. The main difference with the other observables
 analyzed is that here the two active quarks are the light ones. The
 discrepancies might hint at a possible sizeable renormalization of the axial
 coupling for light constituents quarks. On the other hand the value for the 
 ratio $R$ in
 Eq.~(\ref{R}) agrees with the  HQS prediction and with the one evaluated using
 a combination of lattice and experimental data, being also close to a QCDSR
 determination.

\begin{acknowledgments}
We thank J. M. Flynn for useful discussions. This research was supported by DGI and FEDER funds, under contracts
BFM2002-03218, BFM2003-00856 and  FPA2004-05616,  by the Junta de Andaluc\'\i a and
Junta de Castilla y Le\'on under contracts FQM0225 and
SA104/04, and it is part of the EU
integrated infrastructure initiative
Hadron Physics Project under contract number
RII3-CT-2004-506078.  C. Albertus wishes to acknowledge a grant  
from Junta de Andaluc\'\i a. J. M. Verde-Velasco acknowledges a grant
(AP2003-4147) from the
Spanish Ministerio de Educaci\'on y Ciencia.
\end{acknowledgments}

\appendix
\section{Matrix elements for the leptonic decay of pseudo scalars and vector
mesons}
\label{app:lep}
The  matrix element needed for the evaluation of the pseudoscalar decay constant
is given by
\begin{eqnarray}
\hspace{-3cm}\left\langle 0 \left|\, J_{A\, 0}^{f_1\,f_2}(0)
\,\right|\, P, \vec{0}\,
\right\rangle_{NR}
&=& \sqrt3\int\,d^3p\sum_{s_1, s_2}
\hat{\phi}^{(P)}_{(s_1,f_1),\,(s_2,f_2)}(\,\vec{p}\,)
\frac{(-1)^{\frac{1}{2}-s_2}}{(2\pi)^{\frac{3}{2}}
\sqrt{2E_{f_1}(\vec{p}\,)2E_{f_2}(\vec{p}\,)}}\nonumber\\
&&\hspace{3cm}\bar{v}_{s_2,f_2}(\,\vec{p}\,)\, 
\gamma_0\gamma_5\,u_{s_1,f_1}(-\vec{p}\,)
\nonumber\\
&=&\sqrt3\int\,d^3p\sum_{s_1, s_2}
\hat{\phi}^{(P)}_{(s_1,f_1),\,(s_2,f_2)}(\,\vec{p}\,)
\frac{(-1)^{\frac{1}{2}-s_2}}{(2\pi)^{\frac{3}{2}}
\sqrt{2E_{f_1}(\vec{p}\,)2E_{f_2}(\vec{p}\,)}}\nonumber\\
&&\hspace{3cm}\bar{u}_{-s_2,f_2}(\,\vec{p}\,)\, 
\gamma_0\,u_{s_1,f_1}(-\vec{p}\,)\nonumber\\
&=&i\,\frac{\sqrt3}{\pi}\int_0^\infty d|\vec{p}\,| \
\hat{\Phi}^{(P)}_{f_1,\,f_2}(|\vec{p}\,|)\,|\vec{p}\,|^2
\sqrt{\frac{(E_{f_1}(\vec{p}\,)+m_{f_1})(E_{f_2}(\vec{p}\,)+m_{f_2})}
{4E_{f_1}(\vec{p}\,)E_{f_2}(\vec{p}\,)}}\nonumber\\
&&\hspace{3cm}\left(
1-\frac{|\vec{p}\,|^2}{(E_{f_1}(\vec{p}\,)+m_{f_1})(E_{f_2}(\vec{p}\,)+m_{f_2})}\right)
\end{eqnarray}
where we have used the fact that $v_{s,f}(\,\vec{p}\,)=
\gamma_5\,u_{-s,f}(\,\vec{p}\,)$.\\
Similarly for the vector meson case
\begin{eqnarray}
\hspace{-2cm}\left\langle 0 \left|\, J_{V\, 3}^{f_1\,f_2}(0)
\,\right|\, V,\, 0\, \vec{0}\,
\right\rangle_{NR}
&=& \sqrt3\int\,d^3p\sum_{s_1, s_2}
\hat{\phi}^{(V,\,0)}_{(s_1,f_1),\,(s_2,f_2)}(\,\vec{p}\,)
\frac{(-1)^{\frac{1}{2}-s_2}}{(2\pi)^{\frac{3}{2}}
\sqrt{2E_{f_1}(\vec{p}\,)2E_{f_2}(\vec{p}\,)}}\nonumber\\
&&\hspace{3cm}\bar{v}_{s_2,f_2}(\,\vec{p}\,)\, 
\gamma_3\,u_{s_1,f_1}(-\vec{p}\,)
\nonumber\\
&=&\sqrt3\int\,d^3p\sum_{s_1, s_2}
\hat{\phi}^{(V,0)}_{(s_1,f_1),\,(s_2,f_2)}(\,\vec{p}\,)
\frac{(-1)^{\frac{1}{2}-s_2}}{(2\pi)^{\frac{3}{2}}
\sqrt{2E_{f_1}(\vec{p}\,)2E_{f_2}(\vec{p}\,)}}\nonumber\\
&&\hspace{3cm}\bar{u}_{-s_2,f_2}(\,\vec{p}\,)\, 
\gamma_3\gamma_5\,u_{s_1,f_1}(-\vec{p}\,)\nonumber\\
&=&\frac{-\sqrt3}{\pi}\int_0^\infty d|\vec{p}\,| \
\hat{\Phi}^{(V)}_{f_1,\,f_2}(|\vec{p}\,|)\,|\vec{p}\,|^2
\sqrt{\frac{(E_{f_1}(\vec{p}\,)+m_{f_1})(E_{f_2}(\vec{p}\,)+m_{f_2})}
{4E_{f_1}(\vec{p}\,)E_{f_2}(\vec{p}\,)}}\nonumber\\
&&\hspace{3cm}\left(
1+\frac{|\vec{p}\,|^2}{3(E_{f_1}(\vec{p}\,)+m_{f_1})(E_{f_2}(\vec{p}\,)+m_{f_2})}\right)
\end{eqnarray}

\section{Expression for the $V^\mu(|\vec{q}\,|)$ matrix element}
\label{app:v}
The expression for $V^\mu(|\vec{q}\,|)$ is given by
\begin{eqnarray}
V^\mu(|\vec{q}\,|)&=&\sqrt{2M_B2E_D(|\vec{q}\,|)}\
_{\stackrel{}{NR}}
\left\langle\, D,\,
-|\vec{q}\,|\,\widehat{\vec{k}}\,\bigg|\, (J^{c\,b}_V)^\mu(0)\,
\bigg| \, B,\,\vec{0}\right\rangle_{NR}\nonumber\\
&=&\sqrt{2M_B2E_D(|\vec{q}\,|)}\ \int\,d^3p\sum_{s_1, s_2}
\left(\hat{\phi}^{(D)}_{(s_1,c),\,(s_2,f_2)}(\,\frac{m_{f_2}}{m_c+m_{f_2}}\,
|\vec{q}\,|\widehat{\vec{k}}+
\vec{p}\,)\right)^*\ 
\sum_{s_1^\prime}\hat{\phi}^{(B)}_{(s_1^\prime,b),\,(s_2,f_2)}(\,\vec{p}\,)
\nonumber\\
&&\hspace{3.75cm}\frac{1}{\sqrt{2E_c(|\vec{q}\,|\widehat{\vec{k}}+\vec{p}\,)2E_b(\vec{p}\,)}}
\,\bar{u}_{s_1,c}(-|\vec{q}\,|\,
\widehat{\vec{k}}-\vec{p}\,)\, 
\gamma^\mu\,u_{s_1^\prime,b}(-\vec{p}\,)
\end{eqnarray}
where $f_2$ represents  a light $u$ or $d$ quark. 

Going a little further we have
\begin{eqnarray}
V^0(|\vec{q}\,|)&=&\sqrt{2M_B2E_D(|\vec{q}\,|)}\ \int\,d^3p\ \frac{1}{4\pi}
\left(\hat{\phi}^{(D)}_{c,\,f_2}(|\,\frac{m_{f_2}}{m_c+m_{f_2}}\,
|\vec{q}\,|\widehat{\vec{k}}+
\vec{p}\,|)\right)^*\,
 \hat{\phi}^{(B)}_{b,\,f_2}(|\,\vec{p}\,|)\nonumber\\
&&\hspace{1cm}\sqrt{\frac{\bigg(E_{c}(|\vec{q}\,|\widehat{\vec{k}}+\vec{p}\,)
+m_{c}\bigg)\ 
\bigg(E_{b}(\vec{p}\,)+m_{b}\bigg)}{4E_{c}(|\vec{q}\,|\widehat{\vec{k}}+\vec{p}\,)
\ E_{b}(\vec{p}\,)}}
\left(
1+\frac{|\vec{p}\,|^2 +p_z\,|\vec{q}\,|}{\bigg(E_{c}(|\vec{q}\,|
\widehat{\vec{k}}+\vec{p}\,)+m_{c}\bigg)\ \bigg(E_{b}(\vec{p}\,)+m_{b}\bigg)}\right) 
\end{eqnarray}
In the case of equal masses $m_b=m_c$ and for $|\vec{q}\,|=0\ (w=1)$ we will
obtain
\begin{eqnarray}
\left.V^0(|\vec{q}\,|)\right| _{|\vec{q}\,|=0}&=&2M_B\nonumber\\
\end{eqnarray}
Similarly
\begin{eqnarray}
V^3(|\vec{q}\,|)&=&-\sqrt{2M_B2E_D(|\vec{q}\,|)}\ \int\,d^3p\ \frac{1}{4\pi}
\left(\hat{\phi}^{(D)}_{c,\,f_2}(|\,\frac{m_{f_2}}{m_c+m_{f_2}}\,
|\vec{q}\,|\widehat{\vec{k}}+
\vec{p}\,|)\right)^*\,
 \hat{\phi}^{(B)}_{b,\,f_2}(|\,\vec{p}\,|)\nonumber\\
&&\hspace{1cm}\sqrt{\frac{\bigg(E_{c}(|\vec{q}\,|\widehat{\vec{k}}+\vec{p}\,)
+m_{c}\bigg)\ 
\bigg(E_{b}(\vec{p}\,)+m_{b}\bigg)}{4E_{c}(|\vec{q}\,|\widehat{\vec{k}}+\vec{p}\,)\ E_{b}(\vec{p}\,)}}
\left(\frac{p_z}{E_{b}(\vec{p}\,)+m_{b}}+
\frac{p_z +|\vec{q}\,|}{E_{c}(|\vec{q}\,|
\widehat{\vec{k}}+\vec{p}\,)+m_{c}}
\right) 
\end{eqnarray}

\section{Expressions for the $V^{(*)}_{\lambda,\,\mu}$ and 
$A^{(*)}_{\lambda,\,\mu}$ matrix elements}
\label{app:va}
The expressions for  the $V^{(*)}_{\lambda,\,\mu}$ and 
$A^{(*)}_{\lambda,\,\mu}$ matrix elements are given by
\begin{eqnarray}
V^{(*)}_{\lambda,\,\mu}(|\vec{q}\,|)&=&\sqrt{2M_B2E_{D^*}(|\vec{q}\,|)}\ {}_{\stackrel{}{NR}}
\left\langle\, D^*,\,\lambda\ 
-|\vec{q}\,|\,\widehat{\vec{k}}\,\bigg|\, (J^{c\,b}_V)_\mu(0)\,
\bigg| \, B,\,\vec{0}\right\rangle_{NR}\nonumber\\
&=&\sqrt{2M_B2E_{D^*}(|\vec{q}\,|)}\ \int\,d^3p\sum_{s_1, s_2}
\left(\hat{\phi}^{(D^*,\,\lambda)}_{(s_1,c),\,(s_2,f_2)}(\,
\frac{m_{f_2}}{m_c+m_{f_2}}\,
|\vec{q}\,|\widehat{\vec{k}}+
\vec{p}\,)\right)^*
\ \sum_{s_1^\prime}\hat{\phi}^{(B)}_{(s_1^\prime,b),\,(s_2,f_2)}(\,\vec{p}\,)
\nonumber\\
&&\hspace{3.9cm}\,\frac{1}{\sqrt{2E_c(|\vec{q}\,|\,
\widehat{\vec{k}}+\vec{p}\,)\,2E_b(\vec{p}\,)}}
\,\bar{u}_{s_1,c}(-|\vec{q}\,|\,
\widehat{\vec{k}}-\vec{p}\,)\, 
\gamma_\mu\,u_{s_1^\prime,b}(-\vec{p}\,)\nonumber\\
A^{(*)}_{\lambda,\,\mu}(|\vec{q}\,|)&=&\sqrt{2M_B2E_{D^*}(|\vec{q}\,|)}\ {}_{\stackrel{}{NR}}
\left\langle\, D^*,\,\lambda\ 
-|\vec{q}\,|\,\widehat{\vec{k}}\,\bigg|\, (J^{c\,b}_A)_\mu(0)\,
\bigg| \, B,\,\vec{0}\right\rangle_{NR}\nonumber\\
&=&\sqrt{2M_B2E_{D^*}(|\vec{q}\,|)}\ \int\,d^3p\sum_{s_1, s_2}
\left(\hat{\phi}^{(D^*,\,\lambda)}_{(s_1,c),\,(s_2,f_2)}(\,
\frac{m_{f_2}}{m_c+m_{f_2}}\,
|\vec{q}\,|\widehat{\vec{k}}+
\vec{p}\,)\right)^*
\ \sum_{s_1^\prime}\hat{\phi}^{(B)}_{(s_1^\prime,b),\,(s_2,f_2)}(\,\vec{p}\,)
\nonumber\\
&&\hspace{3.9cm}\,\frac{1}{\sqrt{2E_c(|\vec{q}\,|\,
\widehat{\vec{k}}+\vec{p}\,)\,2E_b(\vec{p}\,)}}
\,\bar{u}_{s_1,c}(-|\vec{q}\,|\,
\widehat{\vec{k}}-\vec{p}\,)\, 
\gamma_\mu\gamma_5\,u_{s_1^\prime,b}(-\vec{p}\,)\nonumber\\
\end{eqnarray}
So that
\begin{eqnarray}
V^{(*)}_{-1,\,2}(|\vec{q}\,|)
&=&-\frac{1}{\sqrt2}\sqrt{2M_B2E_{D^*}(|\vec{q}\,|)}\ \int\,d^3p\ \frac{1}{4\pi}
\left(\hat{\phi}^{(D^*)}_{c,\,f_2}(|\,
\frac{m_{f_2}}{m_c+m_{f_2}}\,
|\vec{q}\,|\widehat{\vec{k}}+
\vec{p}\,|\,)\right)^*
\hat{\phi}^{(B)}_{b,\,f_2}(|\,\vec{p}\,|)\nonumber\\
&&\hspace{1cm}\sqrt{\frac{\bigg(E_{c}(|\vec{q}\,|\widehat{\vec{k}}+\vec{p}\,)
+m_{c}\bigg)\ 
\bigg(E_{b}(\vec{p}\,)+m_{b}\bigg)}{4E_{c}(|\vec{q}\,|\widehat{\vec{k}}+\vec{p}\,)\ E_{b}(\vec{p}\,)}}
\left(
\frac{p_z}{E_{b}(\vec{p}\,)+m_{b}}-
\frac{p_z+|\vec{q}\,|}{E_{c}(|\vec{q}\,|
\widehat{\vec{k}}+\vec{p}\,)+m_{c}}\right)\nonumber\\
\end{eqnarray}%
\begin{eqnarray}
A^{(*)}_{-1,\,1}(|\vec{q}\,|)
&=&-\frac{i}{\sqrt2}\sqrt{2M_B2E_{D^*}(|\vec{q}\,|)}\ \int\,d^3p\ \frac{1}{4\pi}
\left(\hat{\phi}^{(D^*)}_{c,\,f_2}(|\,
\frac{m_{f_2}}{m_c+m_{f_2}}\,
|\vec{q}\,|\widehat{\vec{k}}+
\vec{p}\,|\,)\right)^*
\hat{\phi}^{(B)}_{b,\,f_2}(|\,\vec{p}\,|)\nonumber\\
&&\hspace{1cm}\sqrt{\frac{\bigg(E_{c}(|\vec{q}\,|\widehat{\vec{k}}+\vec{p}\,)
+m_{c}\bigg)\ 
\bigg(E_{b}(\vec{p}\,)+m_{b}\bigg)}{4E_{c}(|\vec{q}\,|\widehat{\vec{k}}+\vec{p}\,)\ E_{b}(\vec{p}\,)}}
\left(1+
\frac{2p_x^2-|\,\vec{p}\,|^2-p_z\,|\vec{q}\,|}{\bigg(E_{c}(|\vec{q}\,|
\widehat{\vec{k}}+\vec{p}\,)+m_{c}\bigg)\,\bigg(E_{b}(\vec{p}\,)+m_{b}\bigg)}
\right)\nonumber\\ 
\end{eqnarray}%
%
\begin{eqnarray}
A^{(*)}_{0,\,0}(|\vec{q}\,|)
&=&-i\sqrt{2M_B2E_{D^*}(|\vec{q}\,|)}\ \int\,d^3p\ \frac{1}{4\pi}
\left(\hat{\phi}^{(D^*)}_{c,\,f_2}(|\,
\frac{m_{f_2}}{m_c+m_{f_2}}\,
|\vec{q}\,|\widehat{\vec{k}}+
\vec{p}\,|\,)\right)^*
\hat{\phi}^{(B)}_{b,\,f_2}(|\,\vec{p}\,|)\nonumber\\
&&\hspace{1cm}\sqrt{\frac{\bigg(E_{c}(|\vec{q}\,|\widehat{\vec{k}}+\vec{p}\,)
+m_{c}\bigg)\ 
\bigg(E_{b}(\vec{p}\,)+m_{b}\bigg)}{4E_{c}(|\vec{q}\,|\widehat{\vec{k}}+\vec{p}\,)\
 E_{b}(\vec{p}\,)}}
\left(\frac{p_z}{E_{b}(\vec{p}\,)+m_{b}}+
\frac{p_z+|\vec{q}\,|}{E_{c}(|\vec{q}\,|
\widehat{\vec{k}}+\vec{p}\,)+m_{c}}\right)\nonumber\\
\end{eqnarray}
\begin{eqnarray}
A^{(*)}_{0,\,3}(|\vec{q}\,|)
&=&-i\sqrt{2M_B2E_{D^*}(|\vec{q}\,|)}\ \int\,d^3p\ \frac{1}{4\pi}
\left(\hat{\phi}^{(D^*)}_{c,\,f_2}(|\,
\frac{m_{f_2}}{m_c+m_{f_2}}\,
|\vec{q}\,|\widehat{\vec{k}}+
\vec{p}\,|\,)\right)^*
\hat{\phi}^{(B)}_{b,\,f_2}(|\,\vec{p}\,|)\nonumber\\
&&\hspace{1cm}\sqrt{\frac{\bigg(E_{c}(|\vec{q}\,|\widehat{\vec{k}}+\vec{p}\,)
+m_{c}\bigg)\ 
\bigg(E_{b}(\vec{p}\,)+m_{b}\bigg)}{4E_{c}(|\vec{q}\,|\widehat{\vec{k}}+\vec{p}\,)\ E_{b}(\vec{p}\,)}}
\left(1+
\frac{2p_z^2-|\,\vec{p}\,|^2+p_z\,|\vec{q}\,|}{\bigg(E_{c}(|\vec{q}\,|
\widehat{\vec{k}}+\vec{p}\,)+m_{c}\bigg)\,\bigg(E_{b}(\vec{p}\,)+m_{b}\bigg)}
\right)
\end{eqnarray}
%


\end{document}